\documentclass[fleqn,usenatbib]{mnras}

\usepackage{newtxtext,newtxmath}
\usepackage{cleveref}
\usepackage{mathrsfs}
\usepackage{verbatim}
\crefformat{section}{\S#2#1#3} % see manual of cleveref, section 8.2.1
\crefformat{subsection}{\S#2#1#3}
\crefformat{subsubsection}{\S#2#1#3}
\usepackage[T1]{fontenc}

\DeclareRobustCommand{\VAN}[3]{#2}
\let\VANthebibliography\thebibliography
\def\thebibliography{\DeclareRobustCommand{\VAN}[3]{##3}\VANthebibliography}

\usepackage{graphicx}	% Including figure files
\usepackage{amsmath}	% Advanced maths commands

\usepackage{amssymb}	% Extra maths symbols
\usepackage{bm}
\title[CFD with CDUGKS]{Computational Fluid Dynamics with the Coupled Discrete Unified Gas Kinetic Scheme (CDUGKS)}

\author[Alvaro Zamora, et al.]{
Alvaro Zamora,$^{1}$\thanks{E-mail: pizza@stanford.edu (AZ)}
Elliott Slaughter,$^{2}$
Tom Abel$^{1}$
\\
$^{1}$Kavli Institute for Particle Astrophysics and Cosmology, Stanford University, SLAC National Accelerator Laboratory, Menlo Park, CA 94025, USA
\\
$^{2}$SLAC National Accelerator Laboratory, Menlo Park, CA 94025, USA
}

\date{Accepted XXX. Received YYY; in original form ZZZ}

\pubyear{2015}

\begin{document}
\label{firstpage}
\pagerange{\pageref{firstpage}--\pageref{lastpage}}
\maketitle

\begin{abstract}
In this paper, we introduce our open source implementation of the Coupled Discrete Unified Gas Kinetic Scheme (CDUGKS), a phase space scheme capable of handling a wide range of flow regimes. We demonstrate its performance on several well known test problems from the astrophysical fluid dynamics literature such as the 1D Sod shock tube and Einfeldt rarefaction, 2D Kelvin-Helmholtz instability, 1D thermoacoustic wave, a triangular Gresho vortex, a sine wave velocity perturbation. For these problems, we show that the code can simulate flows ranging from the inviscid/Eulerian regime to the free-streaming regime, capturing shocks and emergent diffusive processes in the appropriate regimes. We also use a variety of Prandtl numbers to demonstrate the scheme's ability to simulate different thermal conductivities at fixed viscosity. The scheme is second-order accurate in space and time and, unlike many solvers, uses a time step that is independent of the mean free path of the gas. Our code (MP-CDUGKS) is public under a CC0 1.0 Universal license and is available on \href{https://github.com/alvarozamora/CDUGKS}{GitHub}.
\end{abstract}


%%%%%%%%%%%%%%%%%%%%%%%%%%%%%%%%%%%%%%%%%%%%%%%%%%

%%%%%%%%%%%%%%%%% BODY OF PAPER %%%%%%%%%%%%%%%%%%

\section{Introduction}
Computational fluid dynamics has proven to be an indispensable tool in astrophysics. For many decades, equations derived within the Chapman-Enskog framework, including those governing Eulerian hydrodynamical and Navier-Stokes flows, have been employed to model astrophysical fluids \citep[e.g.][]{Flash, Athena, Enzo, Athena++}. These first-order techniques have been applied in flow regimes ranging from dense stellar interiors \citep{DenseStellarInterior_Meakin_2007} to rarefied voids in cosmological simulations \citep{Void_Xu_2016}. Much attention has been dedicated to ensuring these schemes have desirable properties e.g. being stable, conservative, and total-variation diminishing \citep{Toro}. Furthermore, a significant amount of effort has gone into incorporating physics such as radiative transfer \citep[e.g.][]{RadT}, chemistry routines and cooling \citep[e.g.][]{grackle_Smith_2016}, non-ideal equations of state \citep[e.g.][]{nonideal_Colella1985}, and to ensure the schemes robustly capture shocks and hypersonic flows. However, not much effort has been put forth by the astrophysics community in investigating the intrinsic error underlying the equations made by only using perfect or first order linear-gradient constitutive relations, and how such errors manifest themselves in simulations both qualitatively and quantitatively. In this work, we employ a phase space method -- particularly, a gas kinetic scheme named CDUGKS \citep{CDUGKS} -- to identify and demonstrate flow regimes for common test problems where the Eulerian and Navier-Stokes approximations are not good descriptions. In addition, we explore the properties of flows outside of this linear-gradient constitutive regime with several new test problems. Put simply, the aim of this paper is to explore and demonstrate the capabilities of CDUGKS and to introduce our open source, massively parallel implementation, MP-CDUGKS\footnote{\url{ https://github.com/alvarozamora/CDUGKS}}.

The equations of Eulerian hydrodynamics are employed in many astrophysical fluid solvers due to the low computational cost relative to Navier-Stokes or phase space methods. For the equations to be valid, there is the assumption that the mean free path of the collisional matter in any given region of a simulation is much smaller than any length scale that one is interested in. It is at the cost of fixing the collisionality that this inviscid approximation produces a computationally feasible model of fluids that is used in e.g. Enzo \citep{Enzo}. Incorporating a linear-gradient constitutive relation (Navier-Stokes), which recovers the ability to tune the collisionality in the low Knudsen number regime, may cause a significant slowdown due to the requirement to resolve the length scale of diffusive processes (e.g. viscosity and conduction). Some workarounds, such as the STS of \cite{STSS} found in Athena++ \citep{Athena++}, can get around this constraint via an approximation that results in a significant speed-up for larger viscosities. However, the accuracy of the Navier-Stokes approximation itself breaks down for larger viscosities. The gas kinetic scheme CDUGKS recovers the ability to simulate a gas of any collisionality at a large (but fixed) computational cost by resolving the higher-dimensional phase space.

There has been a growing interest in gas kinetic schemes based on the BGK model \citep{BGK} in the last decade, and there now exists many different iterations of schemes based on this BGK operator in the literature \citep[e.g.][]{GKS_TAN2018214, DUGKS_Guo_2015, UGKS_MHD, CDUGKS, relativisticBGKVlasovMaxwell}. These include schemes with electromagnetism and relativity. The BGK collision operator appears in the PDE of the full distribution function $f(\bm x, \bm \xi, t)$ as

\begin{equation}
    \label{eq:BGK}
    \frac{\partial f}{\partial t} + \bm{\xi}\cdot\nabla f + \bm{a}\cdot \nabla_\xi f =\Omega_\text{BGK}= \frac{f_\text{eq}-f}{\tau},
\end{equation}
where $\bm{\xi}$ is the phase space velocity vector, $\bm{a}$ is the acceleration vector (due to an external potential, e.g. gravity), $f_\text{eq}$ is the distribution in equilibrium, and $\tau$ is a relaxation time. The BGK model has known deficiencies, such as having a fixed Prandtl number (equal to unity), and so the Ellipsoidal Statistical model (ES-Model) and the Shakhov model (S-Model) were developed to allow for flows with $Pr\neq1$ \citep{ES_S_BGK}. In this paper, we focus on CDUGKS, a double distribution function (DDF) scheme which tracks a  velocity distribution function as well as an energy distribution function. This scheme in particular was chosen because, according to \citet{CDUGKS}, it has the following properties:
\begin{itemize}
    \item It is an asymptotic-preserving (AP) scheme. CDUGKS recovers the NS/Euler solution when the relaxation time is taken to be small compared to the relevant timescales to the problem (i.e. small/vanishing Knudsen number) while also recovering the collisionless solution when the relaxation time is much larger than the relevant timescales to the problem (i.e. large Knudsen number).
    \item By using a total energy double-distribution function (TEDDF) model, the scheme is found to be numerically stable compared to those that use an internal energy distribution function (IEDF), which require the calculation of the interfacial heat flux. Unlike some other models which produce negative values for the distribution functions, the use of the TEDDF model guarantees that the distribution functions will be nonnegative.
\end{itemize}

 It is interesting to consider the ways in which viscosity manifests itself in these different fluid models. By construction, Navier-Stokes fluids have a viscosity that operates in a way which constraints the form of the stress tensor terms to be scalar multiples of the velocity gradients, regardless of the magnitude of the viscosity. This results in the intuitive diffusive interpretation of viscosity in which increasing the magnitude of the kinematic viscosity $\nu$ increases the diffusive rate of momentum without bound. In these BGK-based models including the CDUGKS used in this paper, the bulk viscosity $\mu$ appears not as a diffusive process, but rather as a mechanism that changes the local relaxation time $\tau = \mu / p $ (with pressure $p$). Increasing the viscosity does not directly increase the rate at which momentum diffuses in CDUGKS. Instead, increasing the viscosity increases the time it would take the local velocity distribution to equilibriate. As we will see in this paper, this results in a maximum momentum diffusion rate for a given set of initial conditions. This is because momentum can only diffuse in a hard-sphere gas if mass that carries momentum diffuses; you cannot have one without the other. For a given set of initial conditions for the phase space of a fluid, the unimpeded collisionless solution with the largest mean free path gives the bounded, maximal diffusion of momentum and energy. Although these two distinct forms of viscosity ostensibly manifest themselves indistinguishably at low Knudsen numbers, they of course produce vastly different results at higher Knudsen numbers.
 
 The paper is organized as follows. First, we discuss the scheme as implemented in our code in \cref{sec:sec_scheme}. We discuss the test problems explored in this paper and the relevant physics within in \cref{sec:sec_problems}. We discuss some implementation details in \cref{sec:discussion}. We conclude our results and discuss potential extensions of the work in \cref{sec:conclusion}

\section{The Coupled Discrete Unified Gas Kinetic Scheme (CDUGKS)} \label{sec:sec_scheme}
For completeness, we present a brief overview of CDUGKS as implemented in our code. For a detailed description of the scheme and an in-depth discussion of its properties, refer to the original work by \citet{CDUGKS} (referred to herein as the CDUGKS paper). For the most part, in this paper we use the same notational conventions as in the CDUGKS paper.

Consider the full distribution function, $f(\bm{x},\bm{\xi}, \bm{\eta}, \bm{\zeta}, t)$, of a 3-dimensional mass continuum constrained to vary in $D$ dimensions at the point in phase space $\bm{x},\bm{\xi}\in\mathbb{R}^D$ and time $t$. For each such point in $\mathbb{R}^D$, the distribution function $f$ captures the mass distribution across each of the $D$ kinetic degrees of freedom $\bm{\xi} = \{\xi_1, \ldots, \xi_D\}$ for the $D$ dimensions allowed to vary. It also captures the mass distribution across the components of $\bm{\eta}$, the $(3-D)$ kinetic degrees of freedom along the dimensions not allowed to vary, as well as across the components of $\bm{\zeta}$, the $K$ internal degrees of freedom. This is the distribution function that obeys equation $\ref{eq:BGK}$. Defining a velocity distribution function (VDF) $g$ and total energy distribution function (TEDF) $b$ as
\begin{gather}
    \label{g}g  = \int f(\bm{x},\bm{\xi}, \bm{\eta}, \bm{\zeta}, t)\ d\bm{\eta}d\bm{\zeta}
    \\ \label{b}b  = \frac{1}{2}\int (\xi^2 + \eta^2 + \zeta^2)f(\bm{x},\bm{\xi}, \bm{\eta}, \bm{\zeta}, t)\ d\bm{\eta}d\bm{\zeta},
\end{gather}
allows us to integrate out the degrees of freedom we are not interested in and results in the respective PDEs
\begin{gather}
    \label{gPDE}
    \frac{\partial g}{\partial t} + \bm{\xi}\cdot\nabla g + \bm{a}\cdot \nabla_\xi g = \frac{g_\text{eq}-g}{\tau_g}
    \\
    \label{bPDE}
    \frac{\partial b}{\partial t} + \bm{\xi}\cdot\nabla b + \bm{a}\cdot \nabla_\xi b = \frac{b_\text{eq}-b}{\tau_b} + \frac{Z}{\tau_{bg}}(g-g_\text{eq}) + g\bm{\xi}\cdot\bm{a},
\end{gather}
for the $D$-dimensional distributions $g$ and $b$, where $\tau_g = \mu/p$, $\tau_b=\tau_g/\text{Pr}$, $\tau_{bg} = \tau_b\tau_g/(\tau_g-\tau_b)$, $Z = \bm{\xi}\cdot\bm{u}-u^2/2$, $\bm{u}$ is the local bulk (average) velocity, $\mu$ is the dynamic viscosity, $p$ is the pressure, and $\text{Pr}$ is the Prandtl number. Note that since the PDEs do not contain momentum and energy diffusion terms, the Prandtl number manifests itself as the ratio of the relaxation times, $\text{Pr} = \tau_g/\tau_b$, instead of the usual $\text{Pr}=\nu/\alpha$ for momentum diffusivity $\nu$ and thermal diffusivity $\alpha$. In this paper, we only consider ideal gases with $p = \rho RT$, where $\rho$ is the mass density, $R$ is the gas constant (equal to 0.5 throughout this paper), and $T$ is the temperature. The equilibrium distributions are given by
\begin{gather}
    \label{geq}
    g_\text{eq} = \frac{\rho}{(2\pi RT)^{D/2}}\exp\bigg(-\frac{(\bm{\xi}-\bm{u})^2}{2RT}\bigg)
    \\
    \label{beq}
    b_\text{eq} = \frac{\xi^2 + (3-D+K)RT}{2}g_\text{eq}.
\end{gather}
As in the CDUGKS paper, a temperature dependent model of viscosity is employed, taking the form
\begin{equation}
    \label{eq:visc}
    \mu = \mu_{r} \Big(\frac{T}{T_{r}}\Big)^w
\end{equation}
where $w$ can be set depending on the fluid model, e.g. 0.5 for a hard-sphere model as adopted in this paper \citep{MathematicalTheoryofGases}. After specifying a reference viscosity $\mu_{r}$ at a reference temperature $T_{r}$, one can determine the viscosity at any point by evaluating \eqref{eq:visc}. 

The local density of mass, momentum, and energy can be found using $g$ and $b$ as
\begin{align}
    &\rho = \int g\ d\bm\xi\\ 
    &\rho\bm u = \int \bm u g d\bm\xi\\ 
    &\rho E = \int b d\bm\xi.
\end{align}
With the adiabatic index 
\begin{equation}
    \gamma = \frac{K+5}{K+3}
\end{equation} the temperature is calculated as 
\begin{equation}
    T = \frac{\gamma - 1}{R}\Big(E-\frac{1}{2}u^2\Big),
\end{equation}
and the energy density as 
\begin{equation}
    \rho E = \frac{1}{2}\rho \bm u^2 + \rho \epsilon 
\end{equation}
with the internal energy $\epsilon = c_v T$ and $c_v = (3+K)R/2$ the specific heat capacity at constant volume.

CDUGKS solves the equations for the fields $\phi = g, b$ by first discretizing the velocity space into a set of discrete velocities $\{\bm\xi_i\}$ and obtaining for the PDE's
\begin{align}
    \label{eq:phiPDE}
    \frac{\partial\phi_i}{\partial t} + \bm \xi _i \cdot \nabla\phi_i = \Omega_{\phi,i} + S_{\phi,i}
\end{align}
for the fields $\phi = g,b$ at each $\bm \xi _i$, where the the collision operators for each $\phi$ are defined as 
\begin{align}
    \Omega_g = \frac{g_\text{eq}-g}{\tau_g}, \\ \Omega_b = \frac{b_\text{eq}-b}{\tau_b}
\end{align}
with source terms
\begin{align}
    \label{eq:gsource}
    &S_g = -\bm a\cdot \nabla_{\bm\xi} g ,\\
    \label{eq:bsource}
    &S_b = \frac{Z}{\tau_{bg}}(g - g_\text{eq}) + \bm a \cdot g\bm\xi - \bm a \cdot \nabla_{\bm\xi} b.
\end{align}
Note that the energy distribution source term $S_b$ is still nonzero in the case of no external acceleration field ($\bm a = 0$) if the Prandtl number is not equal to unity. This is how and when it varies from the original one-distribution BGK model. CDUGKS aims to solve the PDEs for the fields $\phi= g, b$ (\eqref{gPDE} and \eqref{bPDE}) accurately to second order in space and time. It does so via the use of a trapezoid rule for the temporal integration of the collision terms $\Omega_g$ and $\Omega_b$, the use of the midpoint rule for the temporal integration of the advection (flux) term $F$, and by performing piecewise upwind linear spatial reconstruction of the reduced distribution functions. As presented in the CDUGKS paper, a rectangle rule is used for the temporal integration of the source terms $S_g$ and $S_b$. Using these integration rules, CDUGKS uses the finite volume method which discretizes the fluid into spatial cells and uses the cell average 
\begin{equation}
    \phi_{j,i}^n = \frac{1}{|V_j|}\int_{V_j}\phi(\bm x, \bm \xi_i, t_n)\  d\bm{x}
\end{equation}
where $|V_j|$ is the volume of cell $V_j$ to obtain for the update rule
\begin{equation}
    \label{eq:implicit}
    \phi^{n+1}_{j,i} = \phi^n_{j+1}-\frac{\Delta t}{|V_j|}F^{n+1/2}_{\phi,j,i} + \frac{\Delta t}{2}\big[\Omega^{n+1}_{j,i} + \Omega^n_{j,i}\big] + \Delta t S^n_{j,i}
\end{equation}
with the respective collision operator and source term for that particular $\phi$. The microflux $F^{n+1/2}_{\phi,j,i}$ is given by the sum of the interfacial fluxes
\begin{align}
    \label{eq:flux}
    F^{n+1/2}_{\phi,j,i}=\sum_k \bm\xi_i \cdot \bm A^k_j \phi(\bm x^k_j, \bm \xi_i, t_{n+1/2})
\end{align}
where $\bm x^k_j$ is the center of the $k$th face of cell j with outward-facing normal vector $\bm A^k_k$ with area $|\bm A^k_j|$. Here, $\phi(\bm x^k_j, \bm \xi_i, t_{n+1/2})$ is the value of the distribution function at the center of the $k$-th face at $t_{n+1/2}$, which is obtained using piecewise upwind linear interpolation with a Van Leer limiter and a quarter-step leapfrog scheme to integrate to the half-step. Note that the term "upwind" here does not refer to a condition on the bulk velocity $\bm u_j$ at cell $j$, but rather on the specific discrete velocity $\bm\xi_i$. In particular, piecewise upwind linear interpolation refers to interpolating from the $j$-th cell center when $\bm\xi_i \cdot \bm A^k_j > 0$ and from the appropriate adjacent cell center of the interface when $\bm\xi_i \cdot \bm A^k_j < 0$. 

Note that \eqref{eq:implicit} is an implicit update rule. The rule is made explicit by updating the conserved variables $\bm W$ in addition to $g$ and $b$ so that one can compute the equilibrium distributions $\phi_\text{eq}$ and relaxation times $\tau_\phi$ at $t_{n+1/2}$ that are required for the collision terms $\Omega_\phi$ at $t_{n+1}$. To obtain the half-step interfacial flux, a quarter-step leapfrog integration scheme is used. This scheme uses two auxiliary fields to mediate the calculation. One field is one quarter-step forward and the other one quarter-step backward from the current time in regards to the collisional and source terms (but not in the advection):
\begin{align}
\label{eq:phibarplus}&\bar{\phi}^+ = \phi +\frac{h}{2}\Omega + \frac{h}{2}S = \frac{2\tau-h}{2\tau}\phi + \frac{h}{2\tau}\phi_\text{eq} + \frac{h}{2}S \\ 
    \label{eq:phibar}&\bar{\phi}\ \  = \phi -\frac{h}{2}\Omega - \frac{h}{2}S = \frac{2\tau+h}{2\tau}\phi - \frac{h}{2\tau}\phi_\text{eq} - \frac{h}{2}S
\end{align}
where $h = \Delta t/2$ and $\Omega$, $S$ are the corresponding $\Omega_\phi$, $S_\phi$, for $\phi =g, b$. One can imagine these distributions as freezing the current distribution $\phi$ in place at $\bm x$ and allowing only the particles at that location $\bm x$ to interact and lag or advance by a quarter-step in the collisional and external processes with a first-order Euler integration scheme with $\delta t= \pm h/2=\pm \Delta t/4$. These fields are used in integrating \eqref{eq:phiPDE} along a characteristic line terminating at the point $\bm x$ to second order using the midpoint rule. In doing so, one obtains the convenient relation
\begin{equation}
    \label{eq:relation}
    \bar\phi (\bm x, \bm\xi_i, t_n + h) = \bar\phi^+(\bm x - h\bm\xi_i, \bm\xi_i, t_n). 
\end{equation}
that the scheme will exploit. In short, these two alternative distributions were constructed to facilitate integration along this characteristic line. In other words: to first order, the velocity or energy distribution $\bar\phi$ at $\bm x$ (which is one quarter step behind of $\phi$ in the collisional and external processes) is the same as the velocity or energy distribution $\bar\phi^+$ (a quarter-step ahead of $\phi$) one half-time step ago at $\bm x - h \xi_i$. The spatial discrepancy arises because these two alternative distributions $\bar\phi,\ \bar\phi^+$ only differ from $\phi$ due the collisional and external processes and not by the advection. Since we are tracking only a set of discrete velocities $\{\bm \xi_i\}$, we know that the mass $\bar\phi$ tracks at $\bm x$ with velocity $\bm\xi_i$ would be the same as the mass $\bar\phi^+$ tracks at $\bm x - h\bm\xi_i$ one half step ago.

The conserved variables $\bm W_i= (\rho, \rho\bm u, \rho E)$ for the cell located at $\bm x_i$ are given by
\begin{align}
    &\rho= \sum_k w_k g(\bm x_i,\bm\xi_k) \\ &\rho \bm u  = \sum_k w_k \bm\xi_k g(\bm x_i,\bm\xi_k) \\ &\rho E  = \sum_k w_k b(\bm x_i,\bm\xi_k)
\end{align}
where $w_k$ are integration weights (e.g. Newton-Cotes) used to numerically integrate over the velocity space. Because the collision operator is conservative, one can also compute the conserved quantities using $\bar\phi$ sequentially as
\begin{align}
    \label{eq:rhobar}
    &\rho= \sum_k w_k \bar g(\bm x_i,\bm\xi_k) \\ \label{eq:rhoubar} &\rho \bm u  = \sum_k w_k \bm\xi_k \bar g(\bm x_i,\bm\xi_k) + \frac{h}{2}\rho\bm a \\ \label{eq:rhoEbar}&\rho E  = \sum_k w_k \bar b(\bm x_i,\bm\xi_k) + \frac{h}{2}\rho\bm u \cdot \bm a.
\end{align}
This allows for the calculation of the conserved variables at the cell interfaces at $t_{n+1/2}$, as is done in \cref{sec:subsec_scheme}. We now have all of the components of the algorithm.
\subsection{The Algorithm} \label{sec:subsec_scheme}
 Given some cell average state $\phi^n_{j,i}$ at time $t_n$ for velocity $\bm\xi_i$, the scheme starts by computing the corresponding quarter-step advanced state $\bar \phi^{+\ n}_{j,i}$ \eqref{eq:phibarplus}. This is the state that undergoes piecewise upwind linear reconstruction. To perform the reconstruction, the gradients $\bm\sigma_{j,i}$ of $\phi$ are needed. In MP-CDUGKS, a Van Leer limiter is applied as
 \begin{equation}
     \label{eq:VanLeer}
     \sigma = (\text{sgn}(\sigma_1) + \text{sgn}(\sigma_2)) \frac{|\sigma_1||\sigma_2|}{|\sigma_1| + |\sigma_2|} 
 \end{equation}
where $\sigma_1$ and $\sigma_2$ are the backward finite difference derivatives on each side of the interface. The scheme also requires piecewise upwind linear reconstruction of the gradients, which we call $\bm\sigma_{k,j,i}$ -- the gradient vector at the $k$-th face of cell $j$ for each discrete velocity $\bm\xi_i$. For that interpolation, the gradients of each component of the gradients are needed, which are computed in a similar fashion as the gradients themselves.

With the interface value of $\bar\phi^+$ and its gradients, the scheme computes
 \begin{align*}
     \bar\phi^+(\bm x_{k,j} - h\bm\xi_i, \bm\xi_i, t_n) = \bar\phi^+(\bm x_{k,j}, \bm\xi_i, t_n) + (\bm x_{k,j}-\bm x_j)\cdot\bm\sigma_{k,j,i}
 \end{align*}
 at $\bm x_{k,j}$, the $k$th face of cell $j$. Using the relation (\ref{eq:relation}) we know that the left-hand side is equal to $\bar\phi(\bm x_{k,j}, \bm\xi_i, t_{n+1/2})$. 
 
 Now that $\bar\phi$ is obtained at the cell interface one half-step later, one can compute the conserved variables $\bm W_i$ using (\ref{eq:rhobar})-(\ref{eq:rhoEbar}) which is needed to compute $\tau_\phi$ at the boundary to compute $\phi_\text{eq}$ at the boundary using \eqref{geq}, \eqref{beq}, and by inverting \eqref{eq:phibar}. Once $\phi_\text{eq}$ is obtained for all cell faces, the flux can be computed using \eqref{eq:flux}.  
 
 After having computed the source terms using \eqref{eq:gsource} and \eqref{eq:bsource}, which were needed in the very first step to compute $\bar\phi^+$, the scheme finally updates the conserved variables via
 \begin{align}
 \label{eq:rhoupdate}
     & \rho^{n+1}_j = \rho^n_j + \frac{\Delta t}{|V_j|}\int  F^{n+1/2}_{g,j,i}\ d\bm\xi + \Delta t \int S^n_{g,j}\ d \xi
     \\
      \label{eq:rhouupdate}
    & \rho\bm u^{n+1}_j = \rho\bm u^n_j + \frac{\Delta t}{|V_j|}\int  \bm\xi F^{n+1/2}_{g,j,i}\ d\bm\xi + \Delta t \int\bm\xi S^n_{g,j}\ d \xi
    \\
     \label{eq:rhoEupdate}
    & \rho E^{n+1}_j = \rho E^n_j + \frac{\Delta t}{|V_j|}\int  F^{n+1/2}_{b,j,i}\ d\bm\xi + \Delta t \int S^n_{b,j}\ d \xi
 \end{align}
 where we note $\rho$ and $\rho\bm u$ are updated using the microflux $F_g$ for the VDF $g$ and $\rho E$ is updated using the microflux $F_b$ for the TEDF $b$. Finally, the scheme computes the new $\phi_\text{eq}$ and updates $\phi$ via 
 \begin{equation}
 \begin{split}
  \label{eq:phiupdate}
      \phi^{n+1}_{j,i} = &
      \bigg(1 + \frac{\Delta t}{2\tau^{n+1}_j}\bigg)^{-1}\bigg[\phi^n_{j,i}+\frac{\Delta t}{2}\bigg(\frac{\phi^{n+1}_{\text{eq},j,i}}{\tau^{n+1}_j} + \frac{\phi^{n}_{\text{eq},j,i}-\phi^{n}_{j,i}}{\tau^n_j}\bigg) \\ &- \frac{\Delta t}{|V_j|} F^{n+1/2}_{\phi,j,i} +\Delta t S^n_{\phi,j,i}\bigg].
 \end{split}
\end{equation}
\section{Test Problems} \label{sec:sec_problems}
We use a series of test problems to demonstrate the capability of CDUGKS in modeling fluid flow regimes in and out of the Eulerian and Navier-Stokes regimes. We explore the 1D Sod shock tube and Einfeldt rarefaction problem, the 2D Kelvin-Helmholtz instability, a 2D shearing problem, a 1D thermoacoustic wave, the 2D Gresho vortex, and a 1D sine wave perturbation problem. We compare the results either to their corresponding analytical solutions in the Eulerian or Navier-Stokes regime or to numerical solutions using either the popular astrophysics/cosmology code Enzo \citep{Enzo} or Athena++ \citep{Athena++}. 

We also explore the effect that the Prandtl number has on the Sod shock tube problem, the Kelvin-Helmholtz instability, the 2D shearing problem, and in the thermoacoustic wave. The Prandtl number is not a fundamental physical parameter. It is typically defined as the ratio of two physical parameters, namely the ratio of momentum diffusivity $\nu$ to energy diffusivity $\alpha$ as $Pr = \nu/\alpha$. The Prandtl number can vary by changing either one of the two parameters, or by changing both. In this paper, the effect the Prandtl number has on fluid dynamics is illustrated by keeping the reference viscosity $\mu_r$ fixed so as to only change the thermal diffusivity of the system \footnote{Recall that there is no explicit diffusive/Laplacian term in the equations governing CDUGKS and, as such, that the Pr used in the scheme is actually the ratio of the relaxation times of the VDF and the TEDF.}.

Each of these test problems are hard coded and can be run by giving the code a different test problem ID at compile time.

\subsection{Sod Shock Tube}
The Sod shock tube problem is a popular test problem for astrophysical fluid codes for three main reasons. Firstly, it is a quick 1D problem with a simple analytical solution in the inviscid regime. Secondly, it tests whether a solver captures shocks and rarefaction properly and may expose a solver's oscillatory or non-conservative behavior. Thirdly, it is a very relevant test problem in astrophysics since shocks are ubiquitous in self-gravitating collisional systems. Much effort has been put into numerical solvers so that they match the analytic shock tube result for strong shocks. However, attempts to match the results of this test problem with high accuracy to the analytic shock tube result are limited in scope to a special case of fluid flow regime i.e. the inviscid regime. This can be an odd benchmark for some solvers, especially particle-based methods which by design use particles with nonzero mean free paths. 

As in the CDUGKS paper, we solve the Sod shock tube problem as a simple validation of the code. The computational domain is $x\in [0,1]$. The initial conditions are $(\rho_1, u_1, P_1) = (1,0,1) $ for $x\leq0.5$ and  $(\rho_2, u_2, P_2) = (1/8,0,1/10)$ for $x> 0.5$. The problem is solved with Dirichlet boundary conditions. The initial conditions are evolved to code time $t=0.15$, with the following simulation parameters:
\begin{itemize}
    \item Two internal degrees of freedom ($K=2$), implying $\gamma = 7/5$.
    \item A viscosity exponent of $w=0.5$, in accordance with the hard-sphere model from kinetic theory.
    \item A Prandtl number $Pr=2/3$.
    \item A spatial resolution of 1024 cells.
    \item A 1D velocity space discretized evenly with $\xi\in [-10, 10]$, resolved with 1025 cells\footnote{The fourth-order Newton-Cotes scheme used requires a size of $4n+1$ for some integer $n\geq 1$.}. 
    \item A CFL safety factor of 0.75.
\end{itemize}
%  \subsubsection{Sod Shock Tube Results}
 
\begin{figure*}
    \centering
    \includegraphics[width=\textwidth]{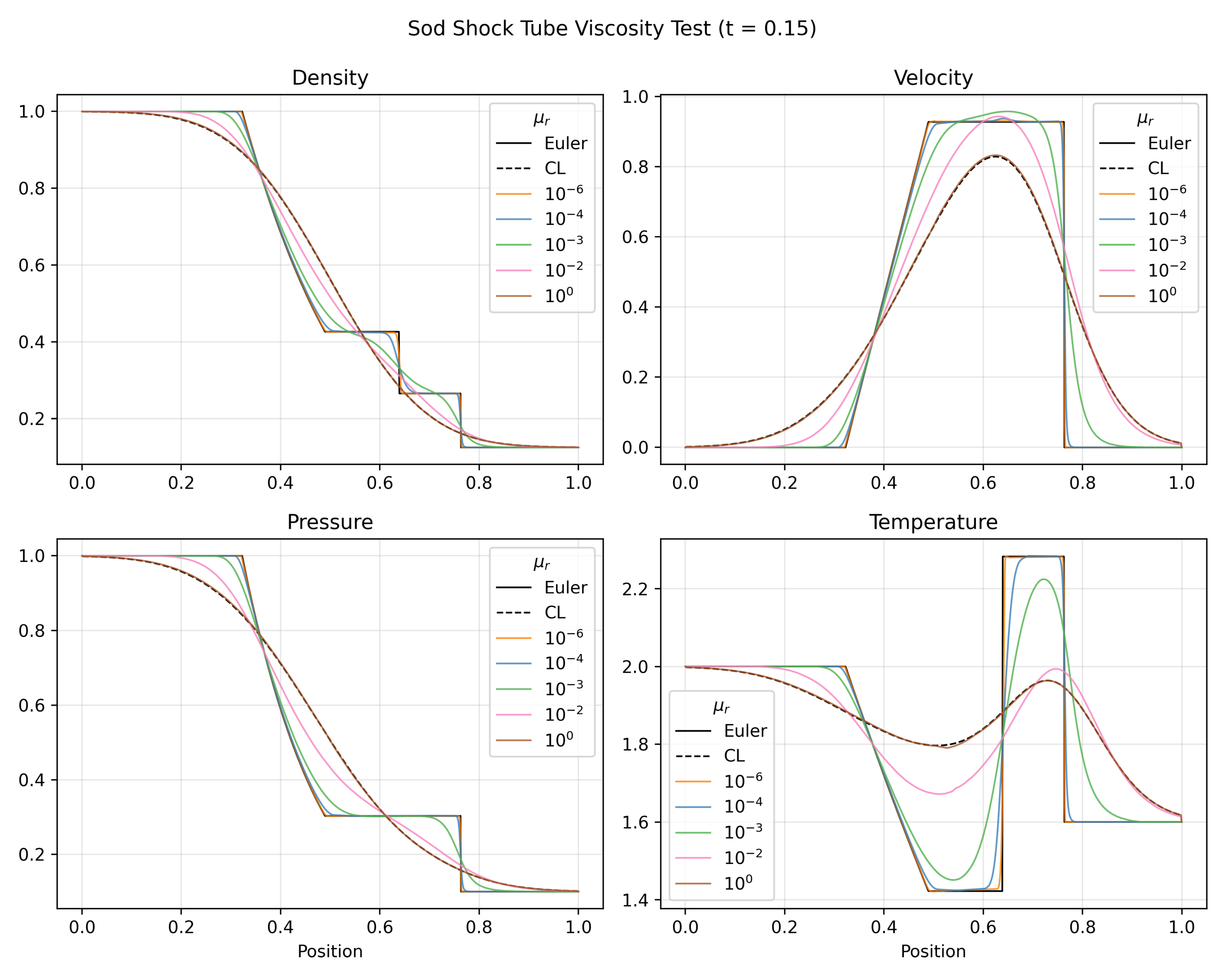}
    \caption{The Sod Shock Tube. Inviscid (Euler) and Collisionless (CL) analytic results are shown along with results from CDUGKS in varying collisional regimes, including the near-Eulerian ($\mu_r=10^{-6}$), Navier-Stokes ($\mu_r=10^{-4}$), transitional ($\mu_r=10^{-3}, 10^{-2}$), and the free-streaming regime ($\mu_r=1$).}
    \label{fig:sod_visc_results}
\end{figure*}

 The versatility of CDUGKS become very clear when considering a problem such as the Sod shock tube problem. Its asymptotic-preserving property allows us to recover the well-known analytic results in the inviscid regime and the collisionless regime, in addition to obtaining results in the Navier-Stokes and transitional regime, all with the same code and for the same computational cost. Figure \ref{fig:sod_visc_results} shows the simulations results for the five runs with $\mu_r = 10^0, 10^{-2}, 10^{-3}, 10^{-4}, 10^{-6}$ at $T_r = 1$ along with the analytic inviscid (Euler) and the collisionless (CL) result. 
 
 Notably, the run with negligible viscosity $\mu_r=10^{-6}$ matches the analytic Eulerian result fairly well. Similarly, the run with very large viscosity $\mu_r = 1$ matches the analytic collisionless result fairly well. For the latter run, a reference viscosity $\mu_r=1$ (mapping to a $Kn \sim 3.2$) for a hard-sphere model was chosen to illustrate the point brought up about NS vs CDUGKS viscosity. That is, there are infinite orders of magnitude above this value in parameter space for this (inverse) collisionality parameter, yet the results are very close to the collisionless case. Increasing the viscosity in CDUGKS to achieve $Kn\sim 10, 100$ may bring the result closer to the collisionless (CL) result, but increasing the viscosity effectively doesn't increase the momentum diffusion rate any more. In the Navier-Stokes model for fluids with conduction, the rate of momentum or energy diffusion would increase along with this parameter without bound (in addition to the growth rate of the respective diffusion length scales). One must be mindful about whether the microphysics in the problem considered allows for such phenomena.

\begin{figure*}
    \centering
    \includegraphics[width=\textwidth]{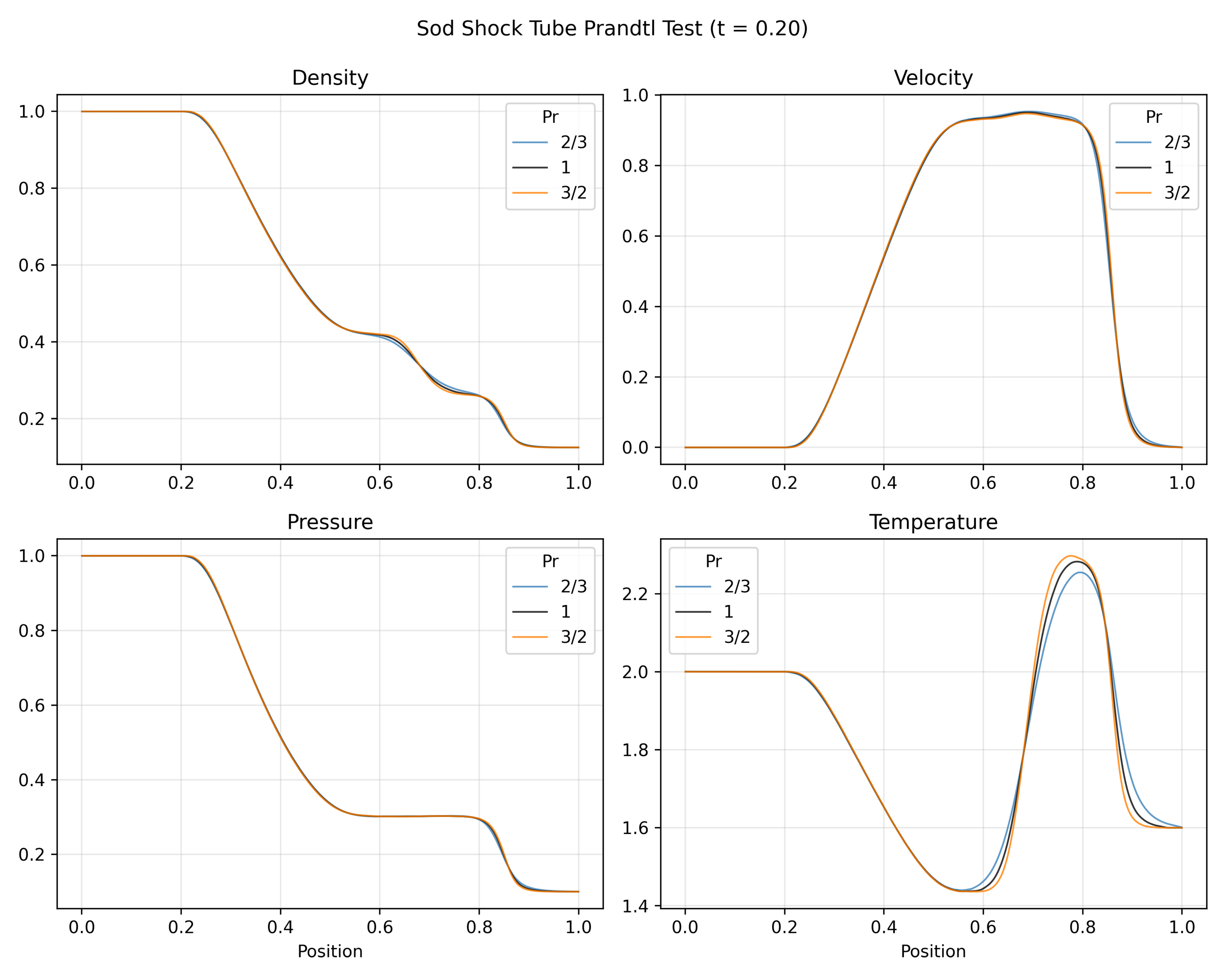}
    \caption{The Sod Shock Tube. Results from CDUGKS at Pr$=2/3, 1, 3/2$ (fixed reference viscosity $\mu_r=10^{-3}$) are shown. As expected, the results vary most in the temperature field.}
    \label{fig:sod_Pr_results}
\end{figure*}

Since the Prandtl number is formally the ratio of the momentum diffusivity to the thermal diffusivity, the effects that the Pr has on the fluid dynamics (for values close to unity) are largest in the viscous Navier-Stokes and transitional regime. To understand why, consider a fluid in the Eulerian regime. Changing the Pr from 2/3 to 3/2 while holding the viscosity fixed changes the thermal diffusivity from negligible to even more negligible. A similar situation arises in the large Knudsen limit. More precisely, this is due the relaxation timescales $\tau_g$, $\tau_b$ being irrelevant if they are much larger or much smaller than the problem-specific hydrodynamical timescale (e.g. sound-crossing time). If they are too small, then the distributions in question are effectively always in equilibrium; if too large, neither distribution relaxes much toward equilibrium at any point in space. In this paper, we demonstrate the effects of the Prandtl number for the hard-sphere model. Figure \ref{fig:sod_Pr_results} shows the results for the Sod shock tube problem with reference viscosity $\mu_r=10^{-3}$ for three different runs: Pr$=2/3, 1, 3/2$.

\subsection{Einfeldt Rarefaction}
\begin{figure*}
    \centering
    \includegraphics[width=\textwidth]{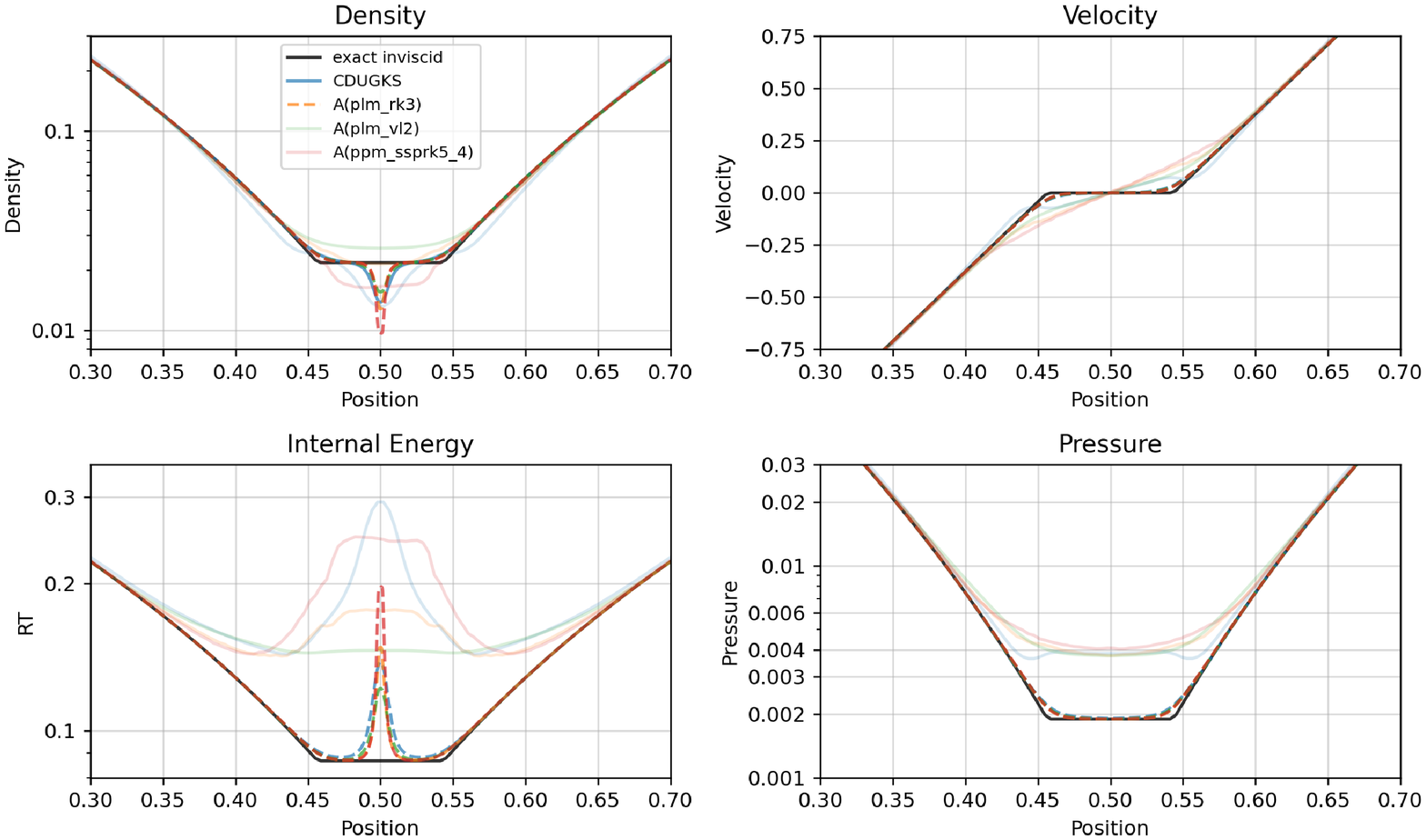}
    \caption{CDUGKS vs Athena++ on the Einfeldt 1-2-0-3 Rarefaction problem at t=0.125. For the Athena++ runs, several different reconstruction methods (PLM, PPM) and integrators (VL2, RK3, SSPRK5\_4) of different order are used. The faint solid lines denote the runs with 256 spatial cells, while the dashed lines denote the runs with 8192 spatial cells.}
    \label{fig:einfeldt}
\end{figure*}

Similar to the Sod shock tube problem, this problem contains a strong discontinuity which typically requires special handling in traditional hydrodynamic methods (i.e. specialized or exact solvers, aggressive slope limiters) to preserve physical values for the conserved quantities. We showcase the capability of CDUGKS in handling strong discontinuities by simulating the 1-2-0-3 Einfeldt Rarefaction problem \citep{Einfeldt, fireball} in the hydrodynamic regime ($\mu_r = 10^{-6}$). The 1-2-0-3 problem is given by a constant density $\rho=1$ and total energy $\rho E = 3$, with a strongly divergent and discontinuous velocity profile $v = 2\ \text{sgn}(x-x_0)$ --- we use the domain $[0,1]$ with $x_0 = \frac{1}{2}$. In addition, we use for the CDUGKS parameters:
\begin{itemize}
    \item Two internal degrees of freedom ($K=2$), implying $\gamma = 7/5$.
    \item A viscosity exponent of $w=0.5$, in accordance with the hard-sphere model from kinetic theory.
    \item A Prandtl number $Pr=2/3$.
    \item A 1D velocity space discretized evenly with $\xi\in [-10, 10]$, resolved with 129 cells.
    \item A CFL safety factor of 0.40
\end{itemize}
We compare results to Athena++ using the 1-2-0-3 problem in the example suite with kinematic viscosity $\nu = 10^{-6}$ and thermal diffusivity $\kappa = \frac{3}{2}10^{-6}$ and with an equal number of spatial cells\footnote{Because Athena++ handles these diffusive processes differently than CDUGKS, these parameters are not physically equivalent but result in diffusive effects that are initially on the same order of magnitude.}. Figure \ref{fig:einfeldt} shows the results of CDUGKS and Athena++ at two different spatial resolutions: 8192 and 256 spatial cells. The dip in the density at the center relative to the inviscid result is the result of the spike in internal energy caused by the strong velocity gradient; to reach pressure equilibrium, the hot gas must rarefy. All solvers of all orders produce results that vary by over 10 percent, highlighting the difficulty of numerical treatment of this problem. Finally, we note that CDUGKS largely agrees with the results of Athena++ despite only using a velocity grid with 129 points. The high resolution Athena++ runs took about 1 minute to run on a single core on an AMD EPYC 7502P (varied across the different integrators and schemes), while the high resolution CDUGKS run took about 2.5 hours using 32 cores on the same processor. Note that this is roughly the Athena++ run time multiplied by the size of the velocity grid; the number of operations in the update rule for a single spatial cell scales linearly with the size of the velocity grid.

 \subsection{Kelvin-Helmholtz Instability}
 The Kelvin-Helmholtz Instability (KHI) is a classic problem which demonstrates the rapid growth of vorticity that arises from small velocity perturbations in shearing flows of different densities. In the inviscid limit and in the absence of gravity, such a perturbation (of any wavelength) grows exponentially at early times. In the Navier-Stokes regime, there exists some minimum perturbation wavelength for which there is rapid growth -- any perturbation on a scale smaller than this wavelength is dampened due to viscosity\footnote{One could say this is the low Reynolds number regime, although we caution against the use of the number since the traditional definition is not Galilean-invariant outside of the pipe setting} \citep{KHImud, KHINSDampeningPaper}. 
 
In this work, a sinusoidal perturbation to the vertical velocity with a wavelength equal to the box size $\lambda = 1$ is chosen, with the following simulation parameters:
\begin{itemize}
    \item Two internal degrees of freedom ($K=2$).
    \item A viscosity exponent of $w=0.5$, in accordance with the simple hard-sphere model from kinetic theory.
    \item A 2D velocity space discretized evenly with each component $\xi_i\in [-10, 10]$, resolved with 101 cells.
    \item A spatial resolution of $100^2$ cells.
    \item A CFL safety number of 0.8.
\end{itemize}
A ramp function is used to create smooth initial conditions which have finite difference gradients that converge with increasing resolution. This was done to avoid the well-documented ill effects of using initial conditions with a sharp interface and to have results to compare to since the inviscid result one gets for the sharp interface from Enzo does not converge with increasing resolution \citep{Robertson_2010}. The ramp function used \citep{Robertson_2010} is given by 
\begin{equation}
    R(y) = \frac{1}{1 + e^{-\frac{2(y-0.25)}{\delta}}}\times\frac{1}{1 + e^{-\frac{2(0.75-y)}{\delta}}}
\end{equation}
with $\delta=0.05$. The horizontal velocity is ramped such that $v_x = -\frac{1}{2}v_r +R(y)v_r$ with relative velocity $v_r = 2.0$, and the density is ramped such that $\rho = \rho_0 + R(y)\delta\rho$ with $\rho_0 = 1$, $\delta\rho=1$. The velocity perturbation is given by $v_y = \delta_y \sin(2\pi x)$ with $\delta_y=0.04$. The fluid is initialized at pressure equilibrium, with $P = 2$.

\begin{figure*}
    \centering
    \includegraphics[width=\textwidth]{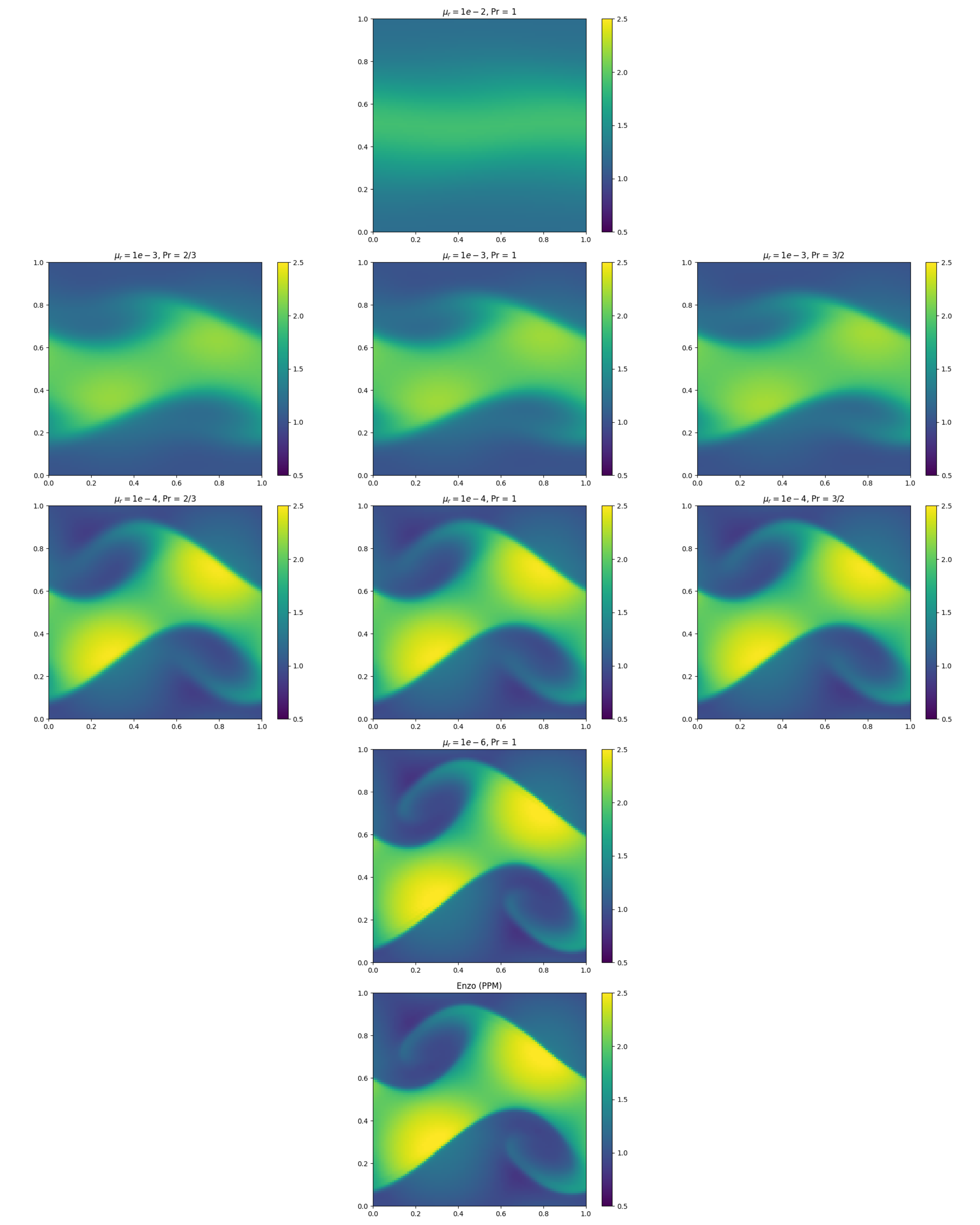}
    \caption{Results (mass density) from CDUGKS and Enzo on the Ramped KHI initial conditions. For the results from CDUGKS, each row corresponds to a different viscosity and each column corresponds to a different Prandtl number. The KHI test problem in Enzo has zero viscosity and zero conductivity, by construction.}
    \label{fig:KHIrho}
\end{figure*}

% \subsubsection{Kelvin-Helmholtz Instability Results}
A total of eight simulations were run, with $\mu_r = 10^{-2}, 10^{-3}, 10^{-4}, 10^{-6}$ at $T_r=1$ with $Pr = 1$. For each of the two $\mu_r = 10^{-3}, 10^{-4}$, two additional simulations with Pr $= 2/3, 3/2$ were run to investigate the effect of the Prandtl Number. The density at $t=1.6$ is presented in Figure \ref{fig:KHIrho}. Each row represents a different $\mu_r$, while each column a different Pr. A ninth simulation was run (in Enzo, at the same spatial resolution) for comparison.

As can be seen in Figure \ref{fig:KHIrho}, the results for the near-inviscid $\mu_r = 10^{-6}$ simulation match the Eulerian result from Enzo very well. In general, increasing the Prandtl number sharpens the result, since the energy diffusion is decreased. As such, and as in the Sod shock tube, the correct Prandtl number for the monatomic gas (Pr$=2/3$) as expected from kinetic theory results in smoother fields. In addition, as the mean free path approaches the box size, the KH mode dampening becomes stronger as seen by the significant diffusion in the run with $\mu_r = 10^{-2}$. Such effects may be of interest in the context of mixing (e.g. heterogeneous vs homogeneous).

\subsection{Shear Problem}\label{sec:shear}

Consider a 2-dimensional periodic square domain $[0,1]^2$ with uniform density and temperature (and thus at pressure equilibrium), and a velocity profile
\begin{equation}
    \bm u_0(x) = (0, \delta u \sin 2\pi x).
\end{equation}
Within the Navier-Stokes framework with nontrivial constant viscosity the evolution of $u_y$ for this problem is well-approximated by ignoring the pressure gradient and momentum flux term, which results in the diffusion equation
\begin{equation}
    \frac{\partial \bm u}{\partial t} = \nu \nabla^2 \bm u
\end{equation}
with solution 
\begin{equation}
    \label{eq:NSshear}
    \bm u (x, t) = \bm (\bm u_0(x)-\Bar{\bm u}) e^{-4\pi^2 \nu t} + \Bar{\bm u}
\end{equation}
where for this problem the spatial average $\Bar{\bm u} = \langle \bm u_0(\bm x) \rangle =0$. The pressure gradient and momentum flux terms give rise to a thermoacoustic wave along the $x$-dimension which, although is a rich problem, shall not be discussed here.

Due to the simplicity of the problem and its emphasis on a non-KHI-exciting shear, it is a great playground for exploring the nature of shear in flows beyond the Navier-Stokes regime. We evolved these initial conditions in CDUGKS up to $t=4.0$ with the following simulations parameters:

\begin{itemize}
    \item Two internal degrees of freedom ($K=2$).
    \item A viscosity exponent of $w=0.0$ (constant viscosity), to be able to compare to analytical approximations.
    \item A Prandtl number of Pr=$2/3$ is used.
    \item A 2D velocity space discretized evenly with each component $\xi_i\in [-6, 6]$, resolved with 129 cells.
    \item A spatial resolution of $128\times1$ cells (there is zero $y$-gradient in any quantity of interest).
    \item A CFL safety number of 0.7.
\end{itemize}
Four simulations were run with $\mu_r = 10^0, 10^{-1}, 10^{-2}, 10^{-3}$.

An interesting and insightful caveat must be discussed before discussing the results. Of course, specifying initial conditions for a phase space method such as CDUGKS requires more than the first few moments of the distribution function; for simplicity, every problem studied in this paper assumes a Maxwell-Boltzmann distribution (equation \ref{geq} and \ref{beq}) initially when given the moments. This initialization and its uniqueness is only appropriate and well-defined in the Eulerian regime. Once given the freedom to deviate from equilibrium, a fluid can manifest an uncountable number of velocity and energy distributions for a given set of the typical 5 moments (mass, momentum, energy). Enforcing that the distribution give rise to a particular stress tensor as in the Navier-Stokes system is one way to constrain and pick one out from the many. One is actually implicitly considering a different set of phase space initial conditions when changing the viscosity parameter in the Navier-Stokes model for the same set of initial moments $\rho$, $\rho v$, $\rho E$.

Inspecting the analytic expectation for the Navier-Stokes result for $v_y$ (equation \ref{eq:NSshear}), we see that the exponential viscous dampening timescale $\tau_{\nu} = (4\pi^2\nu)^{-1} $ shrinks without bound with increasing $\nu$ for fixed initial conditions. Additionally, the analytic approximation should in principle get better with increasing $\nu$, as the viscous timescale becomes a smaller and smaller fraction of the sound-crossing time. In a phase space scheme, the only way momentum can be advected elsewhere is if mass that carries momentum is advected elsewhere. Thus, two things must be true in order to satisfy this analytic NS result for larger and larger viscosity in addition to the conservation laws within CDUGKS. Firstly, one must modify the higher moments of the initial velocity distribution so as to increase the interfacial momentum flux while fixing the temperature and mean momentum. Secondly, it must be true that equations underlying CDUGKS (i.e. equation \ref{eq:BGK}, \ref{gPDE}, \ref{bPDE}) must be able to produce and perpetuate distributions with such moments; it isn't clear if that is indeed the case. In any case, this poses a computational problem since our implementation of CDUGKS uses Newton-Cotes quadrature to integrate a truncated velocity space, which could not feasibly integrate distributions with very high kurtosis or generally larger higher moments.

Alternative initializations for the velocity distributions could be explored with this code. The distribution
\begin{equation}
    \label{eq:NSdist}
    f_1 = f_\text{eq} \bigg( 1 - \frac{1}{n}\sqrt{2RT} \bm{A}\cdot \nabla \ln T - \frac{2}{n} \bm B :\nabla \bm u\bigg)
\end{equation}
for some vector $\bm{A}$ and rank 2 tensor $\bm{B}$ and the equilibrium distribution $f_\text{eq}$ that arises from the first-order Chapman-Enskog expansion \citep{MathematicalTheoryofGases} was not considered for this problem since it allows for negative values for large enough hard-sphere diameter $\sigma$ at only speeds several times the sound speed away from the mean with a high-order approximation. In particular, using the work of \cite{ChapmanThesis} to produce a 150th order accurate expansion for $\mathscr{C}^2\hat{B}(\mathscr{C})$ for values of $\mathscr{C} = |\bm{v} - \bm{u}|/\sqrt{2 R T}\in [0,6]$ for the hard-sphere model, we find that for the shear problem with $\mu_r = 1$, one gets negative values for $f$ for some of the values of $\mathscr{C}\in [0,6]$. This is simply because the values of $\bm{B}$ at fixed $\mathscr{C}$ scale with $\sigma^{-2}$, and so the final term in \eqref{eq:NSdist} diverges for smaller $\sigma$. Thus, as expected, using this truncated distribution from the first-order Chapman-Enskog expansion is nonphysical for the simulations with large mean free paths. With no clearly good alternative for the simulations with larger mean free paths, the Maxwellian-Boltzmann distribution was used for initialization for all runs. 
%\footnote{It is worth noting that, due to the non-Maxwellian initial conditions used by the Navier-Stokes approximation, .}.

\begin{figure}
    \centering
    \includegraphics[width=\columnwidth]{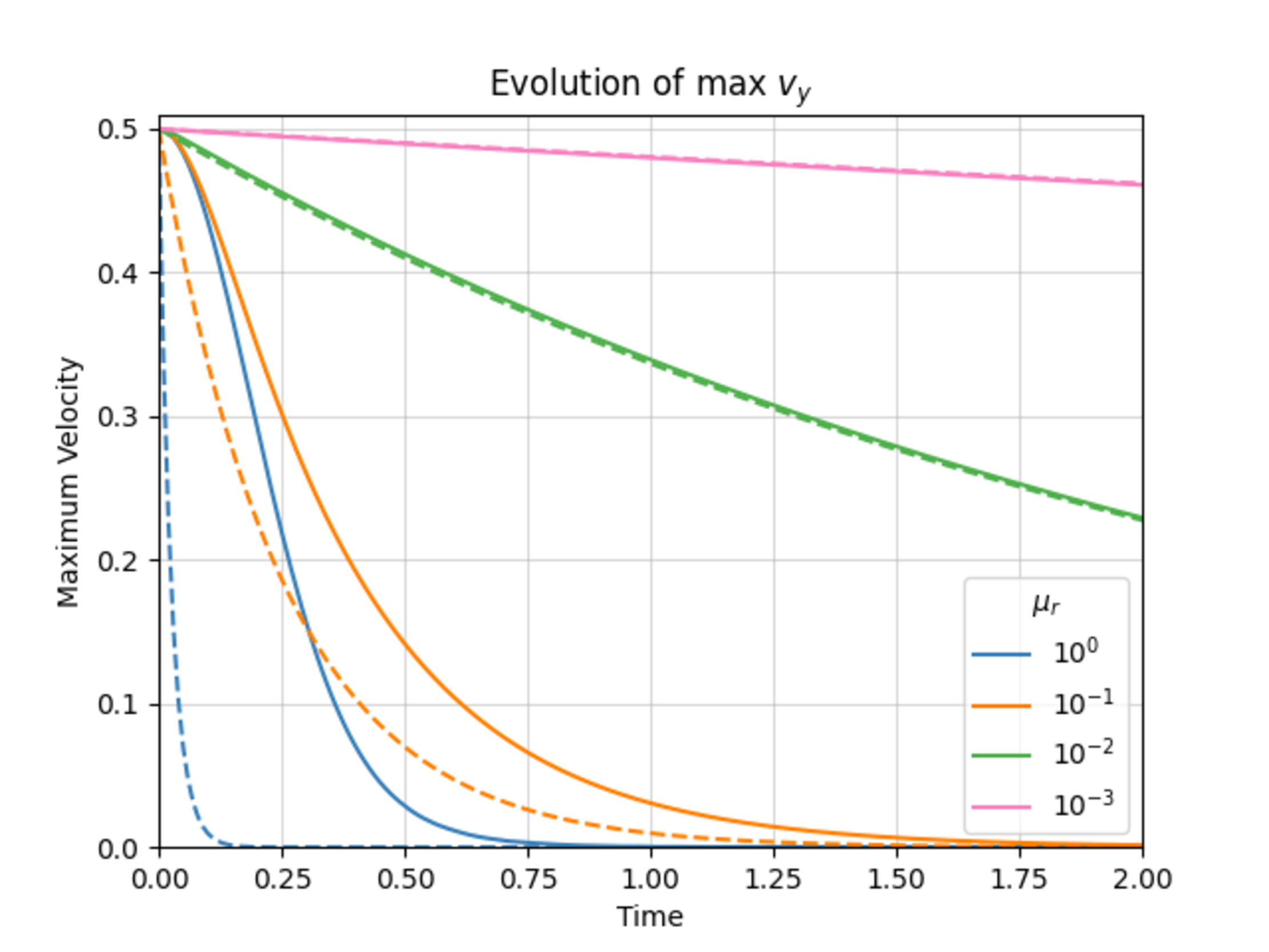}
    \caption{This figure shows how using CDUGKS initialized with a Maxwell-Boltzmann distribution differs from using Navier-Stokes for the shearing box problem. As the viscosity parameter increases, the underlying distribution used by the Navier-Stokes approximation deviates significantly from equilibrium. It becomes clear that choosing initial conditions in this regime is now nontrivial, as one no longer has the uniqueness provided by the Eulerian or Navier-Stokes approximation.}
    \label{fig:shear}
\end{figure}

% \subsubsection{Shear Problem Results}
This problem highlights the ability for CDUGKS to model flows beyond the Navier-Stokes regime, as it shows that we are very far from equilibrium in this region of parameter space. Figure \ref{fig:shear} shows the evolution of the maximum value of $v_y$. As one increases the viscosity, the Navier-Stokes result (here obtained with simulations using Athena++ \citep{Athena++} with equal values for kinematic viscosity and thermal diffusivity, shown with dashed lines) diverges from the CDUGKS result for identical initial conserved macroscopic quantities. If the Navier-Stokes equations were valid for all Kn, our analytic approximation should become better with increasing $\nu$ since the absolute value of the diffusion term would be larger than all other terms in the PDE. However, we see the effect described in the previous section. Namely, without enough mass carrying enough momentum in the tails of the distribution that we track (and without a mechanism for distributions with larger higher moments to emerge), there is simply no manner in which one can match the Navier-Stokes result with a Maxwell-Boltzmann initialization.

An insightful point mentioned previously is that there seems to be a maximum momentum diffusion rate as predicted by CDUGKS (and other BGK-based models). This shows up in the slope of the lines in Figure \ref{fig:shear}; $\frac{d v_\text{y}}{dt}|_{t=0}$ seems to converge to some maximum value with increasing $\mu_r$.

\subsection{Thermoacoustic Wave}
This test problem simply aims to display how conduction is treated in the code since, as for viscosity, there is no explicit diffusion term as there is in Navier-Stokes solvers. The problem begins with an ideal gas at pressure equilibrium, with $\rho(x) = 1 - \delta \sin 4\pi x$, $RT(x) = P_0/\rho$, and $P_0 =1$ with the following simulation parameters:
\begin{itemize}
    \item Two internal degrees of freedom ($K=2$),
    \item A viscosity exponent of $w=0.5$, with $\mu_r = 10^{-3}$.
    \item A 1D velocity space discretized evenly with $\xi\in [-10, 10]$ resolved with 513 cells,
    \item A spatial resolution of $512$ cells,
    \item A CFL safety number of 0.8,
\end{itemize}
the initial conditions were evolved until code time $t=4$ for three simulations with Pr $=2/3, 1, 3/2$, all with $\delta = 0.05$.

\begin{figure*}
    \centering
    \includegraphics[width=\textwidth]{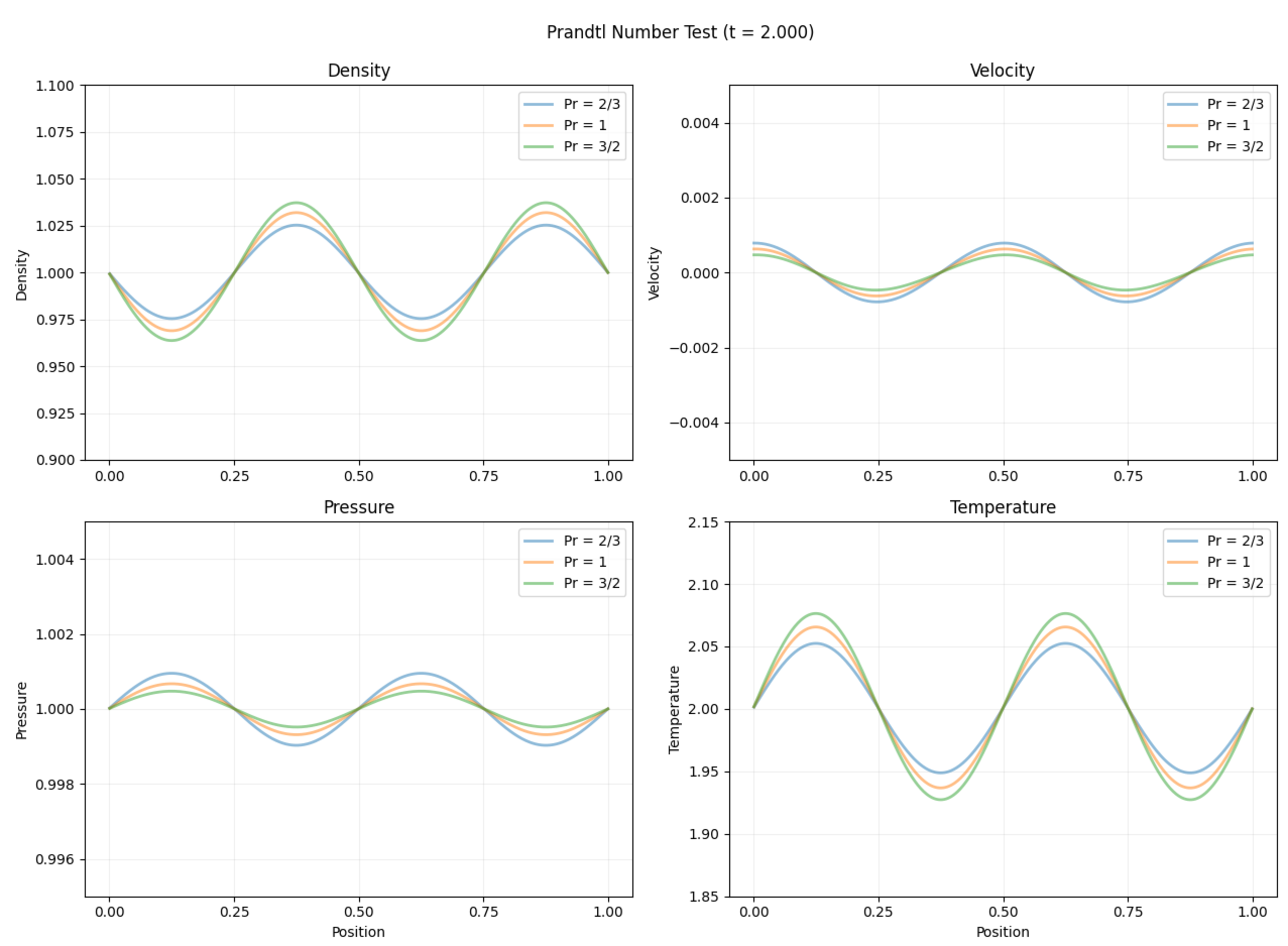}
    \caption{This snapshot shows the difference in the thermoacoustic dynamics with varying Prandtl numbers. The larger Prandtl number simulations exhibit more diffusion in temperature and density, with the latter brought about by stronger pressure and velocity gradients.}
    \label{fig:TA1}
\end{figure*}

\begin{figure*}
    \centering
    \includegraphics[width=\textwidth]{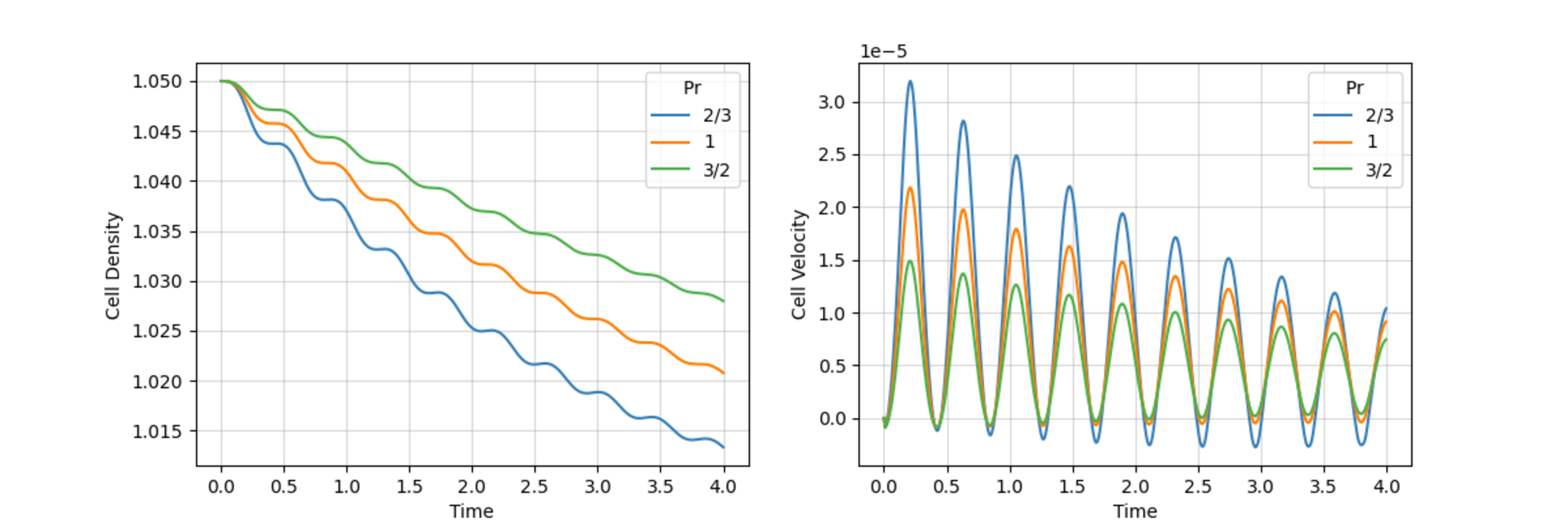}
    \caption{Time evolution of density and velocity in one of the two cells initialized with maximal density $\rho_{\text{max}}$.}
    \label{fig:TA2}
\end{figure*}

% \subsubsection{Thermoacoustic Wave Results}
Figure \ref{fig:TA1} shows a snapshot of the simulation at $t=2$ for the three runs. Here, you can see how in just over approximately one and a half sound crossing times $t_s\sim1.2$ the different Prandtl numbers (thermal diffusivities) bring about different density, velocity and temperature profiles. Since the reference viscosity was fixed, these differences are due entirely to how the energy moves around and relaxes toward equilibrium.
Figure \ref{fig:TA2} shows the time evolution of density and velocity at one of the two cells that are initialized with the maximal density $\rho_{max} = 1 + \delta = 1.05$. Notice how the run with $Pr<1$ has a velocity (in the initially densest cell) which dips well below zero, and the density increases slightly and briefly.

\subsection{Gresho Vortex}
\begin{figure}
    \centering
    \includegraphics[width=\columnwidth]{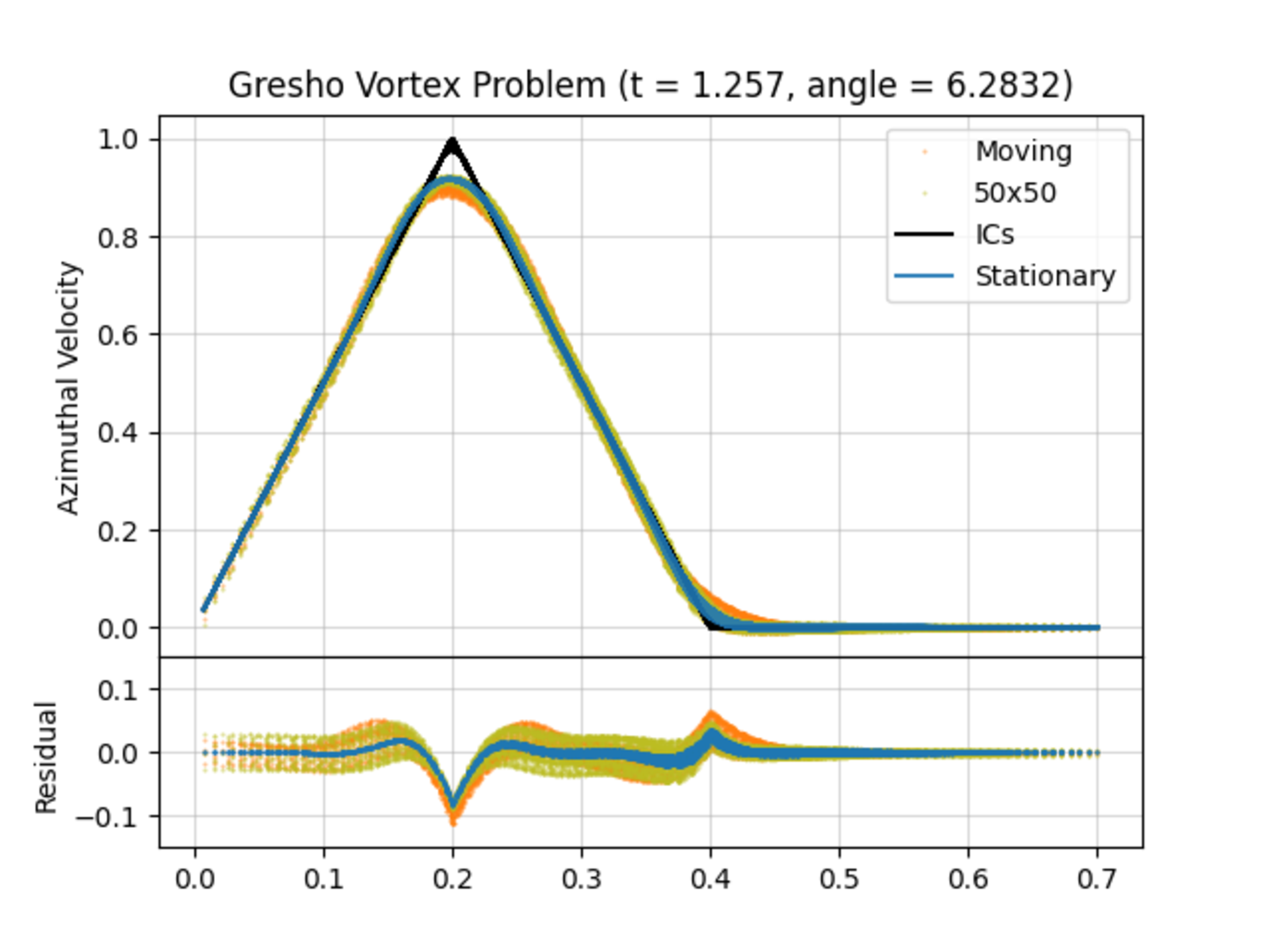}
    \caption{The azimuthal velocity profile for the three simulations, compared to the initial profile. The lower panel shows a scatter plot of the residuals between the final conditions and the initial conditions.}
    \label{fig:GV1}
\end{figure}

The Gresho vortex problem is the cylindrically symmetric cousin of the shear problem covered in \cref{sec:shear}. Thus, here we overlook viscosity and conduction and only analyze the angular momentum conservation properties of CDUGKS in the near-inviscid regime. This is typically a concern for methods that discretize velocities since the discretization (both into a spatial Cartesian grid, but also into a Cartesian grid in velocity space) is not rotationally symmetric for all rotation angles $\phi_R \in (0,2\pi)$, and so some loss of angular momentum is expected during a rotation. The initial conditions are similar to the original triangular vortex \citep{GreshoPaper}, with the azimuthal velocity profile 
\begin{align}
    u_\phi(r) = \begin{cases} 
      5r & r\leq 0.2 \\
      2-5r & 0.2 < r \leq 0.4 \\
      0 & r > 0.4 
   \end{cases}
\end{align}
and complementary pressure profile
\begin{align}
    P(r) = \begin{cases} 
      P_0 + \frac{25}{2}r^2 & r\leq 0.2 \\
     P_0 + \frac{25}{2}r^2 + 4( 1 - 5r - \log 0.2 + \log r) & 0.2 < r \leq 0.4 \\
      P_0 - 2 + 4\log 2 & r > 0.4 
   \end{cases}
\end{align}
with $P_0 = \frac{\rho u_{\text{max}}}{\gamma M^2} = 5$, constant density $\rho = 1$, $\gamma = 7/5$, $M = 1/\sqrt{7}\approx 0.378$, and $u_{\text{max}} = 1$.
Three simulations were run with the following parameters:
\begin{itemize}
    \item Two internal degrees of freedom ($K=2$),
    \item A viscosity exponent of $w=0.5$, with $\mu_r = 10^{-5}$.
    \item A 2D velocity space discretized evenly with $\xi\in [-10, 10]$ resolved with 101 cells,
    \item A spatial resolution of $100\times 100$ cells,
    \item A CFL safety number of 0.8.
\end{itemize}
The first simulation was simply the original problem. The second simulation added a bulk motion $\delta v_x = 3/(2\pi r_\text{max} u_\text{max}) = 15/2\pi$  so that the center of mass would traverse the box three times in the time it takes the fastest ring with $u_\text{max} = 1$ at $r_\text{max} = 0.2$ to complete a rotation. The third simulation used the original initial conditions but used a $51 \times 51$ velocity resolution on the same velocity space domain to explore the resolution dependence of the angular momentum conservation. For each simulation, the initial conditions were evolved for a full rotation of fastest ring (initially at $r = 0.2$), i.e. for a total time $t_\text{sim}=2\pi/5$.
% \subsubsection{Gresho Vortex Results}
The results are well summarized by Figure \ref{fig:GV1}, showing the velocity profiles from all three simulations along with the initial conditions and the residuals from the initial conditions. As expected, both the moving simulation and the simulation with only a $50\times 50$ velocity space resolution are lossier than the stationary simulation. This highlights the need to specify a velocity grid which is fine enough for the problem at hand. The minimum velocity resolution to use should depend on how well angular momentum should be conserved, as well as the dynamic range of the problem.

\subsection{Sine Wave Perturbation}
The sine wave perturbation problem is often explored in the context of the Vlasov-Poisson system. Here we study a similar problem with no gravity. Since this version of CDUGKS uses a representation of the phase space distribution at a fixed, discrete set of velocities, it cannot represent a perfectly cold fluid. Furthermore, it becomes computationally intensive to represent any hypersonic flow since one must be able to resolve the width of the stream in velocity space while also being able to resolve the range of bulk flow velocities present in a simulation. The degree of this computational strain can be quantified by comparing the minimum and maximum characteristic thermal velocity $\sqrt{2 R T}$ to the minimum and maximum residual of the bulk from the mean $\bar{\bm u} - \bm{u(x)}$. Both the minimum and maximum of each quantity have to be compared because similar situations arise for very hot gases, which need larger velocity space domains for some bulk velocity profile to represent all of the mass to machine precision. In any case, in this paper we explore the problem with the cold (finite temperature) initial profile

\begin{align}
    \rho(x) & = \rho_0
    \\ v(x) &= v_0\sin 2\pi x
    \\ E(x) &= C_v T_0 + \frac{1}{2}v(x)^2
\end{align}
with $\rho_0 = 1$, $v_0 = 1$, $T_0 = 0.1$, and $C_v =(3 + K)R/2 = 1$. We evolve these initial conditions forward for reference viscosities $\mu_r$ of $10^{-5}$, $10^{-3}$, $10^{-2}$, $10^{-1}$, and $10^{1}$, roughly representative of the behavior all the way from the Eulerian to the free streaming regime. We use

\begin{itemize}
    \item Two internal degrees of freedom ($K=2$),
    \item A viscosity exponent of $w=0.5$, with $\mu_r = 10^{-5}$.
    \item A 1D velocity space discretized evenly with $\xi\in [-6, 6]$ resolved with 16385 cells,
    \item A spatial resolution of $128$ cells,
    \item A CFL safety number of 0.5.
\end{itemize}

\begin{figure}
    \centering
    \includegraphics[width=\columnwidth]{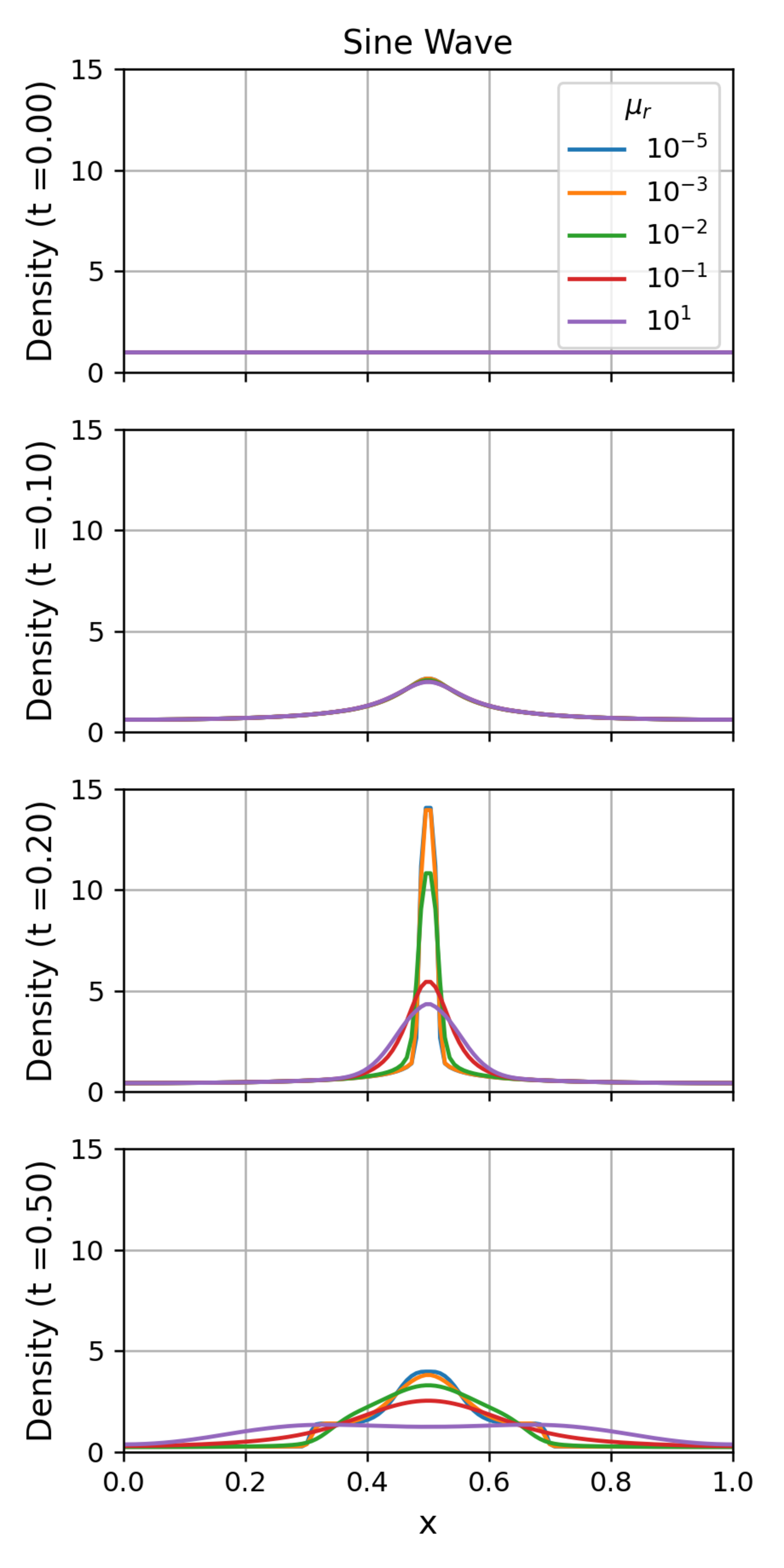}
    \caption{Sine wave perturbation at four different points in time. In order from top to bottom: initial conditions ($t=0$), inflow ($t=0.1$), circa max density ($t=0.2$), outflow (with shocks for the low viscosity regime) ($t=0.5$).}
    \label{fig:SWC_density_evolution}
\end{figure}

% \subsubsection{Sine Wave Perturbation Results}
There are a few temporal points of interest for this problem. The mass distribution converges near the center of the grid to some peak density and then (depending on the fluid regime) can form two shocks that propagate away from the center. Figure \ref{fig:SWC_density_evolution} shows the density evolution at several points in time for the different reference viscosities.

\begin{figure*}
    \centering
    \includegraphics[width=\textwidth]{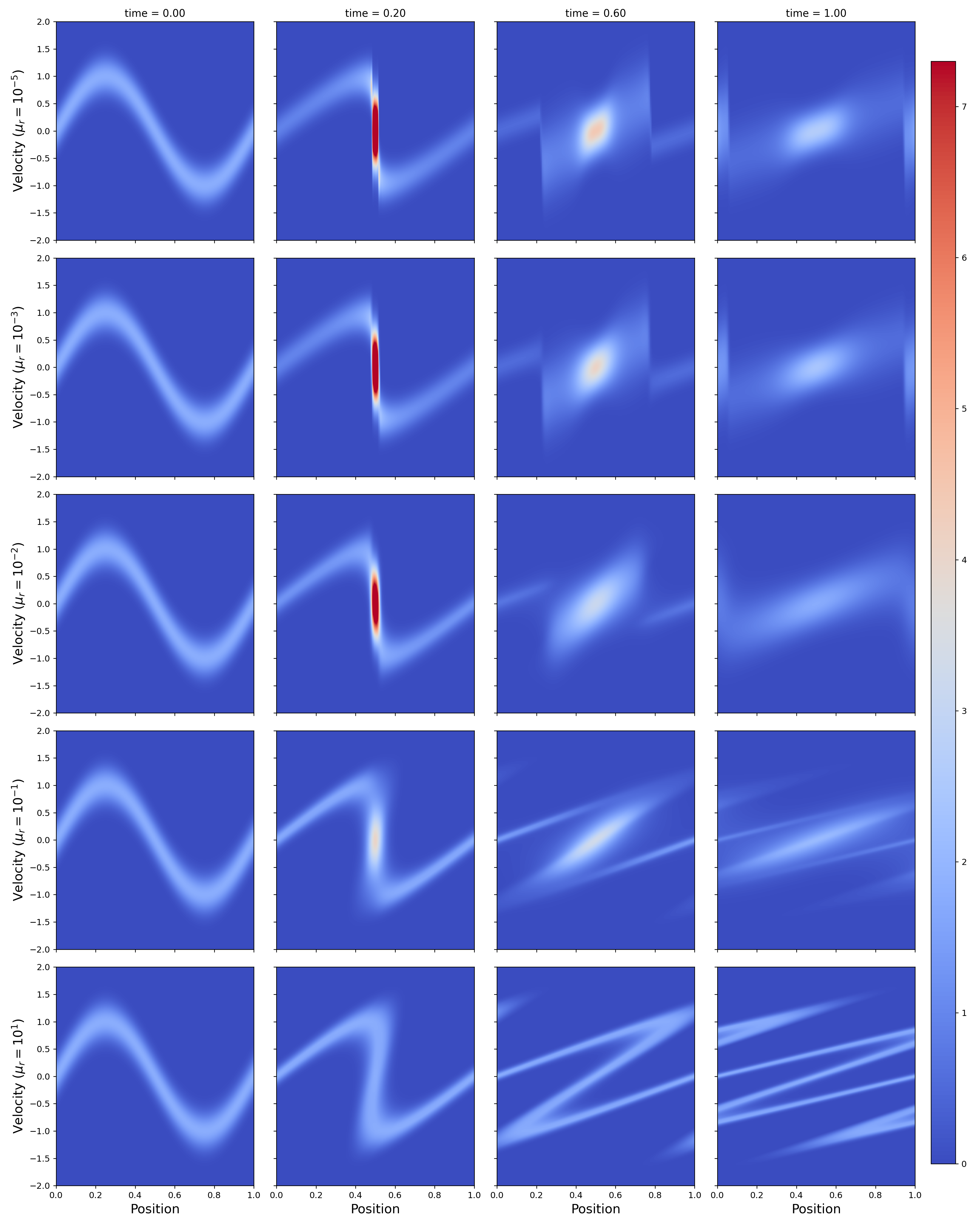}
    \caption{Phase space evolution of the sine wave perturbation at four different points in time. In order from left to right: initial conditions ($t=0$), circa max density ($t=0.2$), outflow (with shock for lower viscosity runs) ($t=0.6$). The final column shows the shock interaction in the Eulerian/NS regime (upper rows), and shows streams overlapping in the free streaming regime (lower rows). To better display the lower densities, the colorbar range here is limited to half the maximum of that in the top row at $t=0.2$.}
    \label{fig:SWC_phase_evolution}
\end{figure*}

It is also quite informative to inspect the phase space evolution (Figure \ref{fig:SWC_phase_evolution}). 
This showcases yet again the ability for CDUGKS to simulate fluids in different regimes at fixed computational cost; the code captures the formation of shocks and non-equilibriated streams (with multi-modal velocity distributions) and propagates them.

\begin{figure*}
    \centering
    \includegraphics[width=\textwidth]{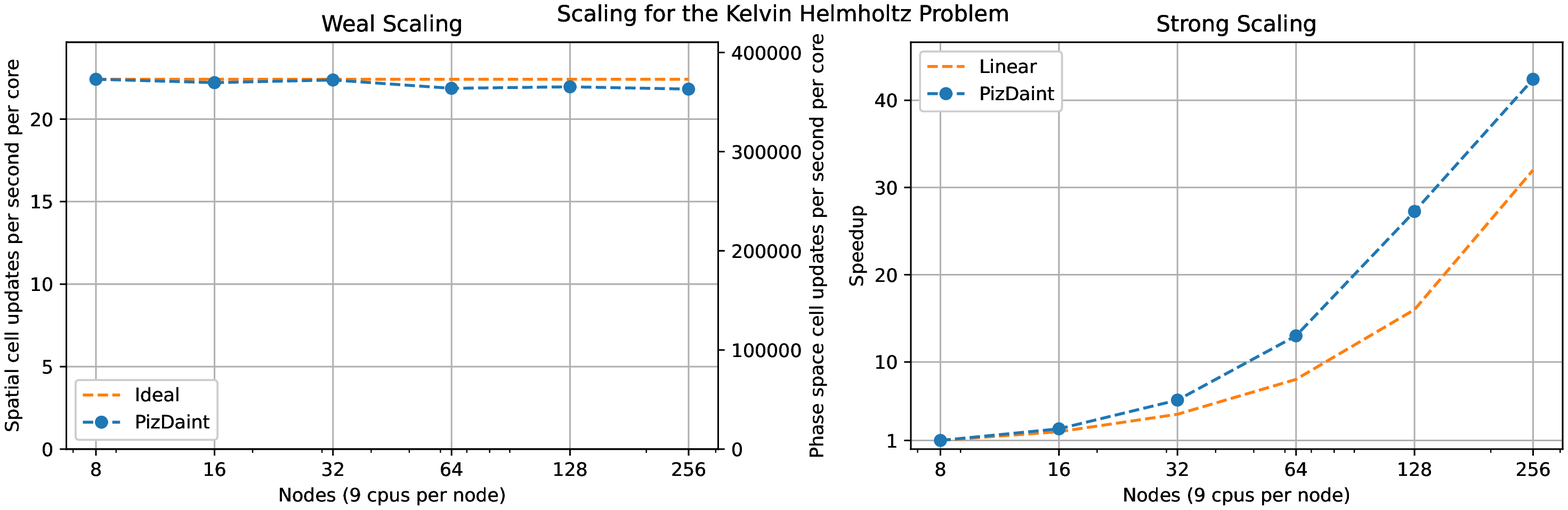}
    \caption{Weak and strong scaling tests on the Kelvin-Helmholtz problem on the Piz Daint cluster. A spatial cell update is defined as an update to all conserved variables in a spatial cell. Likewise, a phase space cell update is defined as an update to both the $g$ and $b$ phase space distributions in a phase space cell. For strong scaling, the baseline (1x speedup) is set at 8 nodes.}
    \label{fig:scaling}
\end{figure*}

\section{Discussion} \label{sec:discussion}
The code was implemented in Regent, a language featuring implicit parallelism written in Legion \citep{Legion}. In addition to the vast physical differences between these phase space methods and the traditional 3D hydro methods used in astrophysics, there are some important algorithmic differences that affect computational performance. Firstly, since CDUGKS uses the discrete velocity method with static grid boundaries, there exists no mass with speeds larger than the largest speed that is being tracked. Thus, the CFL condition for the time step taken becomes
\begin{equation}
    \Delta t = \alpha \frac{\Delta x_\text{min}}{2|\bm\xi_{j,\text{max}}|}
\end{equation}
for $\alpha \in [0,1]$, where $\Delta x_\text{min}$ is the smallest spatial cell width and $|\bm\xi_{j,\text{max}}|$ the largest discrete speed tracked. Note that the original CDUGKS paper has $|\bm\xi_{j,\text{max}} + \bm{u}_{j,\text{max}}|$ in the denominator, but we simply replace the largest average bulk velocity with the largest possible microvelocity to remove the global synchronization required across all nodes. This has its pros and cons. For small problems that e.g. can be run on a single node or laptop, one is limited by smaller time steps which are already smaller than those in usual hydro methods which typically only compare the values of bulk velocities to several local wave speeds (e.g. sound, Alfvén). However, in larger problems that require many nodes to run in a reasonable amount of time, such a CFL condition enables distributed compute without a global synchronization on every time step; with a static spatial and velocity grid, the same time step is used over and over again. 

The calculation of the conserved variables at the boundary, which is required to compute the equilibrium distribution $\phi^{n+1/2}_\text{eq}$ at the boundary, adds a computational strain not seen in typical 3D hydro codes -- this is precisely what makes it a 6D code. For every spatial cell in a $D$-dimensional problem, one must do $D$ numerical integrations over all velocity space for the $D$ right boundaries. This can be thought of as using a local spatial stencil plus a global velocity space stencil. Due to these global velocity space integrals done for every spatial cell, the parallelization is done strictly in spatial subregions. In other words, the grids are not broken up in velocity space. This parallelization strategy will not work for very large problems where the phase space distribution for a single spatial point does not fit on a single node's memory. However it will work for many large problems. 

The weak and strong scaling of the code was evaluated on the Piz Daint cluster in the Swiss National Supercomputing Centre, which uses Intel(R) Xeon(R) CPU E5-2690 v3 (2.60GHz) processors.  For weak scaling tests, we used the Kelvin-Helmholtz test problem with a $2^{2\lfloor N/2\rfloor}\times2^{2\lfloor (N+1)/2\rfloor}$ spatial grid (e.g. $128\times128$ for 8 nodes, $256\times128$ for 16 nodes, $256\times256$ for 32 nodes), and a $129\times129$ velocity grid fixed across all runs. For the strong scaling tests, the same test problem is used with a fixed $256\times256$ spatial grid and $129\times129$ velocity grid. We test the scaling up to up to 256 nodes (a total of 2304 cores) and find that it scales very well. The results are shown in Figure \ref{fig:scaling}. For weak scaling, the throughput per core at 256 nodes is within 3\% of that at 8 nodes. For strong scaling, we find superlinear scaling for the chosen problem up to 256 nodes. The reason it scaled superlinearly was not conclusively determined, but is likely due to better cache utilization. The primary drawback of phase space methods versus traditional hydro methods (the computational cost) can be seen in the large number of phase space cell updates required to get a small number of spatial cell updates. For a full 6D simulation with a $129^3$ spatial grid, the cost per spatial cell update would have been 129 times higher.

\begin{figure*}
    \centering
    \includegraphics[width=\textwidth]{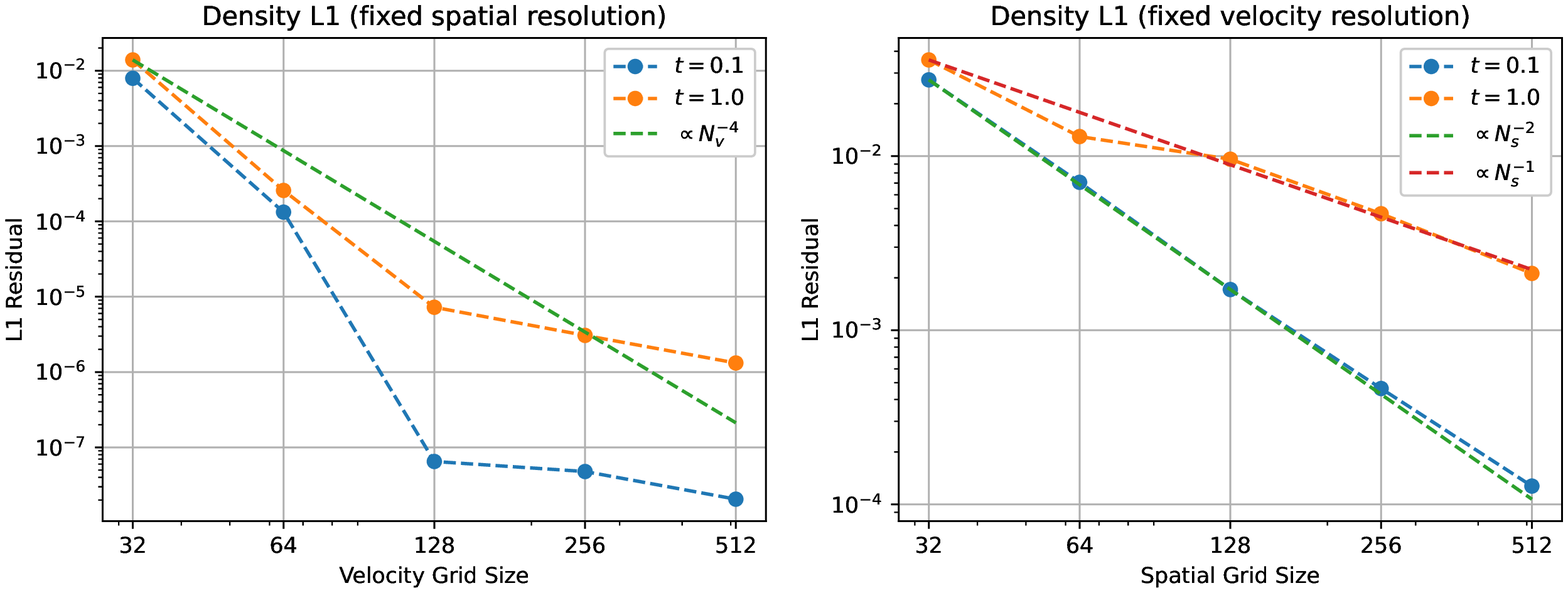}
    \caption{Demonstration of the fourth-order accuracy (velocity grid) and the second-order spatial accuracy of CDUGKS. The Sine Wave Collapse problem is simulated with periodic boundary conditions, and two snapshots are taken at $t=0.1, 1.0$, before and after strong shocks are formed. As expected, the solver reverts to being first-order accurate in space in the presence of strong shocks.}
    \label{fig:norm_convergence}
\end{figure*}

We validate the numerical performance of the code by examining the Sine Wave Collapse problem in the hydrodynamic regime. We begin with the same initial conditions (with temperature $T=1$ instead) and inspect the grid at $t=0.1$. Figure \ref{fig:norm_convergence} shows the L1 norm as a function of velocity grid size (fixed spatial grid $N_s=1024$), and as a function of spatial velocity size (fixed velocity grid $N_v=256$). We see that the algorithm is indeed fourth-order accurate as per the choice of Newton-Cotes weights, and second-order accurate in space. In the presence of strong shocks, solvers that are second-order accurate in space typically revert to being first order accurate. CDUGKS is no exception; we find a $\propto N_s^{-1}$ scaling at $t=1$ when strong shock features form from to the collision of two shocks due to periodic boundary conditions. However, the $N_v^{-4}$ accuracy scaling holds up well for the velocity grid.

Alternative particle-based methods capable of simulating a range of fluid regimes like PolyPIC \citep{PolyPIC} or DSMC \citep{DSMC, AstroDSMC} typically have time steps that depend on parameters other than grid resolution. In addition, due to the nature of discretization, particle-based methods add perturbations at random scales which may grow exponentially and require either a large number of particles or multiple simulations to attenuate statistical noise. On the other hand, since these methods use particles there is no discretization of the velocity space which allows for stricter conservation of angular momentum and intrinsic dynamic resolution of the velocity distribution.

\section{Conclusion and Future Work}\label{sec:conclusion}
In this paper, we introduced our open source code MP-CDUGKS, a massively parallel implementation of CDUGKS. We explored how the code can be used to simulate gases across a wide range of mean free paths, showcasing the asymptotic preserving property, including how it can be used to test the validity of the Eulerian, Navier-Stokes, and free streaming (collisionless) approximations. We discuss how it becomes nontrivial to pick initial conditions for problems outside of the Eulerian/Navier-Stokes regimes, as well as some of the properties of non-equilibrium velocity distributions. We showcase the ability of CDUGKS to simulate gases with different Prandtl numbers, and how it can capture and propagate features such as shocks and even multi-modal velocity distributions (overlapping streams).

There are many ways to extend this work and here we will mention some of them. The first is to use the algorithm to characterize the circumstances in which instabilities grow beyond the Navier-Stokes regime. For the Kelvin-Helmholtz instability, for example, the minimum perturbations wavelengths that grow in the presence of viscosity are known \citep{KHINSDampeningPaper}. However, it is unclear how this relationship changes in the transitional regime just beyond the Navier-Stokes regime. Another extension is to implement a 3D (1 spatial, 2 velocity) version of the code that is able to simulate spherically symmetric problems, so as to study things such as the spherically symmetric gravitational collapse of weakly interacting classical particles. Yet another way is to incorporate more physics into the code (e.g. simple cooling models) to extend the domain of applicability -- algorithms that include electromagnetism and relativity exist \citep[see][]{relativisticBGKVlasovMaxwell}, but there is no open source implementation of them.

Presently in MP-CDUGKS, the velocity grid in a simulation is fixed and should be chosen to be fine enough to resolve the mass in question to within some desired integration error and wide enough to include all of the mass in the system to within the same integration error for the entirety of the simulation. This fixed grid is expensive. However, there are multiple ways one could extend MP-CDUGKS with adaptive velocity grids to increase computational efficiency for problems with high dynamic range \citep[see][]{adaptive1, adaptive2}.

\section*{Acknowledgements}

We are grateful to the support by the Legion team for their help with Regent, Legion, and parallel programming generally. In particular, we thank Alex Aiken for his guidance and support. We further acknowledge the expert support by the Stanford Research Computing Center staff administering and running the Stanford Sherlock HPC Cluster that enabled this work. Finally, we want to thank the Swiss National Supercomputing Centre (CSCS) for enabling the experiments ran on Piz Daint supercomputer through project ID d108.
%%%%%%%%%%%%%%%%%%%%%%%%%%%%%%%%%%%%%%%%%%%%%%%%%%
\section*{Data Availability}

All of the physics problems discussed in this paper are test problems that are available in the source code. The simulation data can be generated easily (even on a personal computer) by running the code with the proper test problem ID. Refer to the code repository for instructions on how to generate the data. Many of the plotting routines are also available in the repository.



\bibliographystyle{mnras}
\bibliography{Paper}

\bsp	% typesetting comment
\label{lastpage}
\end{document}